\documentclass[preprint]{aastex}

\newcommand{\ark}{Akn~564}
\newcommand{\tons}{Ton~S180}
\newcommand{\xsv}{\sigma_{XS}^2}
\newcommand{\fv}{F_{var}}
\newcommand{\fpp}{F_{pp}}
\newcommand{\et}{et al.\ }
\newcommand{\fxv}{\sigma_{XS}^2}

\newcommand{\xmm}{{\it XMM-Newton}}
\newcommand{\xte}{{\it RXTE}}
\newcommand{\euve}{{\it EUVE}}
\newcommand{\chandra}{{\it Chandra}}
\newcommand{\asca}{{\it ASCA}}

\newcommand{\degg}{\hbox{$^\circ$}}
\newcommand{\ls}
{\mathrel{\hbox{\rlap{\hbox{\lower4pt\hbox{$\sim$}}}\hbox{$<$}}}}
\newcommand{\gs}
{\mathrel{\hbox{\rlap{\hbox{\lower4pt\hbox{$\sim$}}}\hbox{$>$}}}}

\begin{document}

\title{X-ray Spectral Variability and Rapid Variability of the Soft X-ray
Spectrum Seyfert 1 Galaxies \ark\ and \tons}
\shorttitle{X-ray Variability of \ark\ and \tons}
\shortauthors{Edelson \et}
\author{ Rick Edelson\altaffilmark{1,2},
	T.\ J.\ Turner\altaffilmark{3,4},
	Ken Pounds\altaffilmark{2},
	Simon Vaughan\altaffilmark{2,5},
	Alex Markowitz\altaffilmark{1},
	Herman Marshall\altaffilmark{6},
	Paul Dobbie\altaffilmark{2},
	Robert Warwick\altaffilmark{2} }

\authoremail{rae@astro.ucla.edu}

\altaffiltext{1}{Astronomy Department; University of California; Los
Angeles, CA 90095-1562; USA}

\altaffiltext{2}{X-ray Astronomy Group; Leicester University; Leicester
LE1 7RH; United Kingdom}

\altaffiltext{3}{Laboratory for High Energy Astrophysics; Code 660;
NASA/Goddard Space Flight Center; Greenbelt, MD 20771; USA}

\altaffiltext{4}{Joint Center for Astrophysics; Physics Department;
University of Maryland Baltimore County; 1000 Hilltop Circle; Baltimore,
MD 21250; USA}

\altaffiltext{5}{Institute of Astronomy; Madingley Road; Cambridge CB3
0HA; United Kingdom}

\altaffiltext{6}{Center for Space Research; Massachusetts Institute of
Technology; 77 Massachusetts Ave.; NE80; Cambridge, MA 02139; USA}

\begin{abstract}

The bright, soft X-ray spectrum Seyfert 1 galaxies \ark\ and \tons\ were
monitored for 35~days and 12~days respectively with \asca\ and \xte\ (and
\euve\ for \tons).
These represent the most intensive X-ray monitoring of any such soft
spectrum Seyfert 1 to date.
Light curves were constructed for \tons\ in six bands spanning 0.1--10~keV
and for \ark\ in five bands spanning 0.7--10~keV.
The short time scale (hours--days) variability patterns were very similar
across energy bands, with no evidence of lags between any of the energy
bands studied.
The fractional variability amplitude was almost independent of energy
band, unlike hard spectrum Seyfert 1s, which show stronger variations in
the softer bands.
It is difficult to simultaneously explain soft Seyferts stronger
variability, softer spectra, and weaker energy-dependence of the
variability relative to hard Seyferts.

There was a trend for soft and hard band light curves of both objects to
diverge on the longest time scales probed ($\sim$weeks), with the hardness
ratio showing a secular change throughout the observations.
This is consistent with the fluctuation power density spectra that showed
relatively greater power on long time scales in the softest bands.
The simplest explanation of all of these is that two continuum emission
components are visible in the X-rays: a relatively hard, rapidly-variable
component that dominates the total spectrum and a slowly-variable soft
excess that only shows up in the lowest energy channels of \asca.
Although it would be natural to identify the latter component with an
accretion disk and the former with a corona surrounding it, a standard
thin disk could not get hot enough to radiate significantly in the \asca\
band, and the observed variability time scales are much too short.
It also appears that the hard component may have a more complex shape than
a pure power-law.

The most rapid factor of 2 flares and dips occurred within $ \sim 1000
$~sec, in \ark\ and a bit more slowly in \tons.
The speed of the luminosity changes rules out viscous or thermal processes
and limits the size of the individual emission regions to $\ls$15
Schwarzschild radii (and probably much less), that is, to either the inner
disk or small regions in a corona.

\end{abstract}

\keywords{galaxies: active --- galaxies: individual (\ark) ---
galaxies: individual (\tons) --- galaxies: Seyfert --- X-rays: galaxies}

\section{ Introduction }
\label{intro}

Seyfert~1 galaxies and quasars are the most powerful sustained, coherent,
quasi-isotropic luminosity sources known, but their distances are so large
that the ``central engines'' in which the luminosity is actually generated
are thought to be orders of magnitude too small to image from Earth.
Therefore, we must rely on indirect probes such as X-ray variability to
infer information about the physical conditions in Seyfert~1s.
This is potentially of general interest, because the luminosity is
ultimately believed to originate in the region of strong gravity ($ \ls 3
R_S $) around a supermassive ($ 10^6 - 10^9 M_\odot $) black hole,
conditions that are unlikely to be reproduced in the laboratory in the
foreseeable future.

Ultraviolet and optical emission-line variability ``reverberation mapping"
studies have yielded key information about the size and structure of the
(much larger) broad-line regions of Seyfert~1s (see Netzer \& Peterson
1997 for a review) that may allow estimation of the mass of the putative
central black hole (Wandel, Peterson \& Malkan 1999).
Although Seyfert~1s are much more strongly variable in the X-rays, less
spectacular results have been seen at those higher energies, quite
possibly because the X-rays probe the smallest size/time scales which may
still lie beyond the limit of current instrumentation.
In recent {\it Advanced Satellite for Cosmology and Astronomy} (\asca) and
{\it Rossi X-ray Timing Explorer} (\xte) surveys, Nandra \et (1997),
Turner \et (1999a) and Markowitz \& Edelson (2001) found evidence that
variations in Seyfert 1s were the largest at softer X-ray energies.
This suggests either that there are two X-ray continuum emission
components, with the softer one showing stronger variability than the
harder one, or that the spectrum of a single component is not
constant, becoming softer as the source brightens.

Evidence for interband lags within the X-rays is less clear cut.
In simultaneous {\it Extreme Ultraviolet Explorer} (\euve), \asca\ and
\xte\ observations of NGC~5548 and MCG--6-30-15, Chiang \et (1999) and
Reynolds (1999) respectively reported evidence that the variations in the
hard X-rays consistently lagged behind those in the soft X-rays by times
shorter than or of order an single spacecraft orbit.
However, Edelson \et (2000) found no such effect in intensive \xte\
monitoring of NGC~3516, and called into question the reality of lag
measurements on time scales shorter than or of order the orbital time
scale.
If such hard interband lags are confirmed, causality arguments would
require rejection of ``reprocessing'' models in which the soft X-rays
are ``secondary'' emission produced by passive reradiation of
``primary'' hard X-ray photons.

Almost all of these studies have involved what could be called ``hard
X-ray spectrum Seyfert~1 galaxies" (or just ``hard Seyferts"): Seyfert 1s
with 2--10~keV power-law slopes in the range $ \Gamma \approx 1.7 - 2.0$.
These sources dominate most X-ray samples, e.g., almost all of the
Piccinotti \et (1982) Seyfert~1s are hard Seyferts.
However, it is now clear that there is a significant population of
Seyfert 1s with much steeper X-ray spectra ($\Gamma \approx 2.1 - 2.6$),
and particularly strong (excess) emission below $\sim$2~keV.
Many of these ``soft X-ray spectrum Seyfert~1 galaxies" (or ``soft
Seyferts") are also optically classified as ``narrow-line" Seyfert~1
galaxies (Osterbrock \& Pogge 1985; Boller, Brandt \& Fink 1996).
However, it is their strongly-variable, steep soft X-ray continua that
really set these objects apart; extreme examples show giant X-ray flares
(as large as a  factor of 100) on time scales of days (e.g., Boller \et
1997).
This rapid X-ray variability also extends to harder X-rays (Turner \et
1999a; Leighly 1999).

The currently favored model is that soft Seyferts are powered by black
holes of  relatively low mass (compared to hard Seyferts of the same
luminosity), accreting at a much higher rate,  closer to the Eddington
limit (Pounds, Done \& Osborne 1995).
In this model the steep X-ray spectrum is a result of enhanced emission
from the putative accretion disk, and the rapid variability results from
the smaller size scales associated with a lower mass black hole (Pounds
\et 2001) and perhaps also an intrinsically less stable accretion
flow.

This paper reports on the most intensive X-ray monitoring of any soft
Seyferts to date: a 35~day simultaneous \asca\ and \xte\ observation of
\ark, and a 12~day simultaneous \asca, \xte\ and \euve\ observation of
\tons.
This paper focuses on the X-ray spectral variability and interband lags in
both objects; other results are reported elsewhere.
The observations and data reduction are reported in the next section,
temporal analyses are performed and discussed in \S~3, the scientific
implications are discussed in \S~4, and a brief summary is given in \S~5.

\section{ Observations and Data Reduction }
\label{ Data }

\subsection{ \ark }
\label{ ark }

\ark\ is the brightest known soft Seyfert in the hard X-ray sky
($ F_{2-10~keV} \approx 2 - 5 \times 10^{-11} $~erg cm$^{-2}$ sec$^{-1}$)
with a steep X-ray spectrum both above $\sim 2$~keV ($\Gamma \approx 2.6
$) and at lower energies (Vaughan \et 1999a; Turner, George \& Netzer
1999b; Pounds \et 2001).
Unfortunately, it has rather large foreground Galactic absorption
($ N_H = 6.4 \times 10^{20} $~cm$^{-2}$; Dickey \& Lockman 1990) that
prevents it from being observed with \euve.
In the observations reported herein, \ark\ was observed simultaneously
with \asca\ over 2000 June 1 -- July 5, with \xte\ over 2000 June 1 --
July 1, surrounded by a total of $\sim$2~yr of \xte\ monitoring once every
$\sim$4.3~day.
Initial results on the \xte\ fluctuation power density spectrum (PDS) and
long/short term variability have been reported in Pounds \et (2001), and
on the \asca\ spectrum in Turner \et (2001b), and other results will be
forthcoming.

\subsubsection{ \asca\ Data }
\label{ ark_asca }

\asca\ has two solid-state imaging spectrometers (SISs; Burke \et 1994)
and two gas imaging spectrometers (GISs; Ohashi \et 1996) yielding data
over an effective bandpass $\sim$0.7--10~keV.
These data were gathered in 1CCD mode.
All the data were screened according to the following criteria: the source
was outside the SAA, the angular offset from the nominal pointing position
was $\leq$ 0.01\degg, the RBM was $\leq$ 500, the cutoff rigidity was
$\leq$ 6 GeV/c, the source was at least 10\degg\ above the Earth's limb
(5\degg\ for the GIS) and at least 20\degg\ from the bright Earth, and the
observations were made $\geq$ 50~s before or after passage through the
terminator.
These are the same methods and screening criteria used by the {\it
Tartarus} (Turner \et 1999a) database.
This resulted in an effective exposure of 1.245~Msec in the GISs,
and 1.109~Msec in the SISs.
Light curves were extracted using source events within extraction cells of
radii 4.8\arcmin\ and 6.6\arcmin\ for the SIS and GIS data, respectively.
In order to increase the signal-to-noise ratio in the light curves, data
from the SIS pair and GIS pair of detectors were (separately) combined,
requiring all time bins to be at least 99\% exposed.
The background was subtracted from these light curves.

\subsubsection{ \xte\ Data }
\label{ ark_xte }

\ark\ was observed once every $\sim$3.2~hr (= 2 orbits) during this
period.
The \xte\ Proportional Counter Array (PCA) consists of five
collimated Proportional Counter Units (PCUs), nominally sensitive to
2--60~keV X-rays (Jahoda \et 1996).
However, only one PCU (number 2) was in use during this campaign.
The present analysis is restricted to the 2--10~keV band, where the PCA is
most sensitive and the systematic errors are best understood.
Data from the top (most sensitive) layer of the PCU array were extracted
using the {\tt REX} reduction script\footnote{See {\tt
http://heasarc.gsfc.nasa.gov/docs/xte/recipes/rex.html}}.
Poor quality data were excluded on the basis of the following acceptance
criteria: the satellite has been out of the South Atlantic Anomaly
(SAA) for at least 20 min; Earth elevation angle $\geq$~10\degg;
offset from optical position of \ark\ $\leq$~0.02\degg; and {\tt
ELECTRON2} $\leq$ 0.1.
This last criterion removes data with high anti-coincidence rate in the
propane layer of the PCA.
These selection criteria typically yielded $\sim$1~ksec good exposure time
per orbit.
The background was estimated using the ``L7--240'' model\footnote{See {\tt
http://lheawww.gsfc.nasa.gov/$\sim$keith/dasmith/rossi2000/index.html}},
which is currently the best available but known to exhibit anomalies
that affect AGN variability studies (e.g., Edelson \& Nandra 1999).
Data were initially extracted with 16 sec time resolution.

\subsection{ \tons }
\label{ tons }

The X-ray spectrum of \tons\ is steep ($\Gamma \approx 2.4$) and, like
\ark\ shows a strong excess at lower energies (Vaughan \et 1999; Turner
\et 2001a).
Although it is not as bright as \ark\  in the hard  X-rays, it does
have a much lower column ($ N_H = 1.5 \times 10^{20} $~cm$^{-2}$;
Stark \et 1992), making possible \euve\ observations.
\tons\ was observed simultaneously for 12~days with \euve, \xte\ and
\asca\
(as well as other telescopes) during 1999 December 3 -- 15.
(The \euve\ and \xte\ observations extended considerably beyond this
period but, for consistency, this paper restricts itself to the 12~day
period during
which all three telescopes were operating.)
Initial results on the \chandra\ spectrum have been reported in Turner \et
(2001a) and the spectral energy distributions will be forthcoming (Romano
\et 2001).

\subsubsection{ \asca\ and \xte\ Data }
\label{ tons_asca }

The \xte\ observations of \tons\ utilized PCUs 0 and 2.
Data were extracted from these as described in \S~2.1.2., the only
differences being in two of the selection criteria.
The {\tt TIME\_SINCE\_SAA} criterion was extended to exclude all data
taken in the 30 minutes following SAA passage.
(The more conservative limit was because \tons\ is fainter than \ark\ and
thus more susceptible to errors in background subtraction.)
An {\tt ELECTRON0} $\le$~0.1 criteria was used to eliminate periods of
high background.

The \asca\ on-source exposures were 327~ksec for the SISs and 396~ksec for
the GISs.
The \asca\ data were reduced and light curves constructed using the same
methods as for Ark~564, except that in this case the predominant SIS
datamode was BRIGHT.

\subsubsection{ \euve\ Data }
\label{ tons_euve }

A light curve was extracted from the \euve\ deep survey (DS) data using
the IRAF subpackage XRAY PROS.
Source counts were summed in a circular aperture of 25 pixels in radius
and the background calculated from a surrounding annulus of 30 pixels in
width.
In some previous analyses of \euve\ DS light curves (e.g., Marshall \et
1996), data with a deadtime-Primbsching correction (DPC) factor $ > 1.25 $
were discarded.
This correction factor accounts for the loss of events due to detector
deadtime and the limited telemetry bandwidth.
As the detector count rate increases, the DPC factor increases and
systematic uncertainties also increase due to incomplete instrument
modeling.
However, during the course of reducing these data it was noted that the DS
DPC factor frequently was above 1.5, significantly greater than the more
typically observed values of 1.0--1.3.
This is most likely due to increased geocoronal emission possibly
associated with the solar maximum and/or decreasing orbital altitude of
\euve.
Data were therefore selected between the more liberal limits of
$ 1.0 < DPC < 2.0 $.
The initial light curve was binned at 50~sec.

\subsection{ Long Time Scale Light Curve Construction }
\label{ light_curves }

The observing logs are given in Table~1.
Essentially identical procedures were used to construct all light curves
for both objects.
Data were extracted in the following subbands: 0.1--0.2~keV (for \euve),
0.7--0.95~keV (for the \asca\ SIS), 0.95--1.3~keV (for the \asca\ SIS
and GIS), 1.3--2~keV (for the \asca\ SIS and GIS), 2--4~keV (for the
\asca\ SIS and GIS and \xte) and 4--10~keV (for the \asca\ SIS and GIS
and \xte).
Data were then binned by the $\sim$95~min orbit (or, in the case of the
\xte\ observations of \ark, every other orbit), and the mean and standard
errors computed.
This yielded light curves with 142--181 points in 12~days for \tons\ (some
were lost due to instrument problems or scheduling conflicts) and 518--520
points in 35~days for \ark\ (231 points for \xte).

\placetable{table1}

Light curves taken with different instruments but in the same bands were
tested for consistency.
In each panel of Figure~1, the data from two different instruments were
plotted in the same graph, after first dividing by the mean.
As the light curves covered the same bands, they should be nearly
identical, modulo the errors, sampling details, and slight mismatches in
energy response.
Confining the analysis first to the \asca\ SIS and GIS data, note that the
light curves show excellent agreement in both the 2--4~keV and 4--10~keV
bands.
(The agreement is similarly good in both sets of softer bands as well.)
This gives confidence in the data and therefore the light curves were
summed to produce a single \asca\ light curve in each band where the GIS
and
SIS overlap, as shown in the second half of Figure~1 and in Tables~2 and
3.

\placefigure{fig1.ps}

\placetable{table2}

\placetable{table3}

Then, these summed data were compared to \xte\ data in the same bands.
Unfortunately, the \xte\ and \asca\ light curves do not show such good
agreement.
Due to its large collecting area, \xte\ is superior to \asca\ for
monitoring the brightest 2--10~keV sources (see, e.g., Edelson \et 2000).
However, \xte\ has a harder spectral response than \asca, so the count
rates are lower for soft Seyferts.
Because \xte\ is also a non-imaging instrument with a high background, the
background must be modeled.
As this estimated background level is larger than the mean count rate for
soft Seyferts but smaller than the mean count rate for many hard Seyferts,
small errors in the background model would thus cause proportionally
larger problems for soft Seyferts.
Indeed, Tables~2 and 3 show that the \xte\ data for both \ark\ and \tons\
have both higher count rates and larger fractional errors in the 4--10~keV
band than in the 2--4~keV band, which would not be expected if only
Poisson statistics contributed to the errors.
Because of this problem, it was decided that the \xte\ data were not
sufficiently reliable for this analysis, and they will not be
scientifically analyzed in this paper.
Instead, the summed \asca\ SIS + GIS data are used, except where the paper
specifically states otherwise (e.g., \S~3.3.).
The resulting light curves are shown in Figure~2.

\placefigure{fig2.ps}

\subsection{ Short Time Scale Light Curve Construction }
\label{ light_curves }

These data were also used to study variations on the shortest accessible
time scales: within a single \asca\ orbit.
These usually lasted 30--40~min without interruption, although a
substantial minority of orbits were affected by SAA passage or minor
telescope problems.

For this purpose, sets of eight 16~sec points were used to measure both
the total 0.7--10~keV count rate and the 2--10~keV/0.7--1.3~keV hardness
ratio.
Standard methods were used to determine the mean and standard error for
each quantity in each 128~sec bin.
\ark\ showed variations of a factor of 2 or larger in 16 orbits.
These data are presented in Figure~3, and will be discussed in \S~3.4.
The largest single-orbit variations seen in \tons\ were four orbits in
which the peak-to-trough variations were 70\%--85\%; these will be
discussed in Romano \et (2001).

\placefigure{fig3.ps}

\section{ Temporal Analysis }
\label{ temporal_analysis }

In the following section the statistical properties of these light curves
are
examined in order to quantify any spectral variability.
A complementary analysis, that of direct spectral fitting to time-resolved
data, was presented in Turner \et (2001).

\subsection{ Long Time Scale Fractional Variability as a Function of
Energy}

The fractional variability amplitude ($\fv$), a common measure of the
intrinsic variability amplitude that corrects for the effects of
measurement noise, is defined as
\begin{equation}
\fv = { 1 \over \langle X \rangle } \sqrt{S^2 - \langle \sigma_{err}^2
\rangle },
\end{equation}
where $S^2$ is the total variance of the light curve, $\langle
\sigma_{err}^2 \rangle $ is the mean error squared and $ \langle X \rangle
$ is the mean count rate (see, e.g., Edelson, Krolik \& Pike 1989).
The error on $\fv$ is
\begin{equation}
\sigma_{\fv} = {1 \over 2 \fv} \sqrt{1 \over N} {S^2 \over \langle X
\rangle^2}
\end{equation}
as discussed in the Appendix.

Tables~2 and 3 the summarize the fractional variability for each
band/instrument, for \ark\ and \tons, respectively.
The fractional variability is also shown as a function of observing energy
in Figure~4.
Note that the variability amplitude is only weakly anticorrelated with
energy.
This is very different from the situation in more ``normal'' hard Seyfert
1s (see references in \S~1), which tend to show stronger variability at
softer X-ray energies.
This will be discussed in \S~4.

Again, note that the \xte\ data show a behavior which is different
than that seen in either of the \asca\ instruments.
The $\fv$s are significantly higher for the \xte\ bands, and in fact for
\tons\ are see to increase with energy.
This apparently spurious \xte\ result was reported (for \ark) by Edelson
(2000a).
Based on the comparison with \asca, we now believe it was almost certainly
due to problems with the \xte\ background.

The \euve\ data on \tons\ appear to show a downturn relative to
extrapolation from the harder \asca\ bands.
However, the \euve\ data are somewhat suspect because they are much
noisier than, e.g., a factor of 4 worse than the \asca\ data, as well as
for reasons given in the next section.

\placefigure{fig4.ps}

\subsection{ Short Time Scale Fractional Variability as a Function of
Energy }
\label{ fpp }

$\fv$ measures the variability power of the total light curve.
As AGN have ``red" PDS (e.g., Edelson \& Nandra 1999), this quantity is
dominated by variations on the longest time scales probed by a given
observation (e.g., Markowitz \& Edelson 2001).
The short time scale variability can be probed by a related parameter,
called the point-to-point fractional variability ($\fpp$), defined as
\begin{equation}
\fpp = { 1 \over \langle X \rangle } \sqrt{ { 1 \over 2(N-1) }
{ \sum_{i=1}^{N-1} ( X_{i+1} - X_i )^2 } - \langle \sigma_{err}^2 \rangle
}
\end{equation}
where $X_i$ is the flux for the $i$th of $N$ orbits.
This measures the variations between adjacent orbits.
This quantity is very similar to the ``Allan Variance"\footnote{See {\tt
http://www.allanstime/AllanVariance/index.html}}.

For white noise, $\fpp$ and $\fv$ give the same value, as we have
confirmed by measuring these quantities for light curves in which the
times have been randomized (to yield a white-noise PDS).
However, for red noise, $\fv$ will be larger than $\fpp$, as the
variations will be larger on longer time scales.
These quantities are tabulated in Tables~2 and 3 and shown in Figure~4.

The \tons\ \euve\ point is formally not defined, as the measured
variability is slightly weaker than just that expected from the errors
alone.
This again suggests that the \euve\ errors are not reliable and the \euve\
$\fv$ and $\fpp$ values should not be taken seriously.

\subsection{ Similarities/Differences between Long and Short Time Scale
Light Curves in Different Bands }
\label{ compare }

The complex nature of the spectral variability of these objects is
concisely illustrated in Figure~5.
Both objects show strong orbit-to-orbit variability in the hardness ratio
($ HR = F_{2-10keV}/F_{0.7-1.3keV}$).
Furthermore, the hardness ratio shows a long-term secular trend for both
objects.
In \ark, it changed over 32~days from $ 0.527 \pm 0.005 $ at the beginning
of the monitoring to $ 0.604 \pm 0.008 $ at the end, and in \tons, it
changed over 9~days from $ 0.588 \pm 0.007 $ at the beginning to $ 0.654
\pm 0.009 $ at the end.
(Mean hardness ratios and standard errors were determined by binning up
hardness ratios in the first and last 3~day periods.)
This is the first time such a clear difference between long and short time
scale variability has been seen in different bands in a Seyfert~1 galaxy.
The implications of this are discussed in \S~4.

\placefigure{fig5.ps}

\subsection{ Rapid Flares and Dips }
\label{ fvar }

It is also interesting to examine the largest and most rapid flux and
spectral flares and dips.
The \ark\ data are more well-suited for this because that source showed
larger variations and the duration of the observation was almost 3 times
that of \tons.
Of the 518 useful orbits in the \ark\ monitoring, 256 have 15 or more
128~sec bins (that is, $\ge$32~min of data).
Of these 256 orbits, 15 (6\%) show peak-to-trough variations of a factor
of $\ge$2 (see Figure~3), and 143 (56\%) show changes of $\ge$50\%.
(The fourth panel in Figure~3 has only 11 points.)
That indicates that the source flux will typically change by a factor of 2
within $\sim$3000~sec, and the fastest factor of 2 variations occur on
very short time scales, $\sim$1000~sec.
In some flares (e.g., the third panel in Figure~3), the source appears to
systematically harden as the flux increases and soften as the flux
declines, in others (e.g., the eighth), it appears to harden as the flux
decreases, and in yet others (e.g., the fifteenth), no clear trend is
apparent.

\subsection{ Fluctuation Power Density Spectra }
\label{ PDS }

In order to further compare the long and short time scale variations in
different energy bands, PDS were measured for each \asca\ energy band.
The \euve\ data were not used because of the large fraction ($>20$\%) of
orbits without data.
The \xte\ data were also not used in this paper for reasons mentioned
earlier.
However, Pounds \et (2001) have already used the full $\sim$2~years of
data on \ark\ to determine the 2--10~keV PDS over a much broader range of
time scales by the technique of Edelson \& Nandra (1999).

The PDS in this paper were derived using standard methods (Oppenheim \&
Shafer 1975, Brillinger 1981), after first creating an evenly-sampled
light curve by interpolating over the few missing points (2\%--3\% of the
data).
A Welch window was applied.
The zero-power and next two (very noisy) lowest-frequency points of each
PDS were ignored and the remaining points binned every factor of 1.8
(0.25 in the logarithm).
The PDS covered a useable frequency range of 1.94 and 1.49 decades
for \ark\ and \tons\, respectively.
Power-law models were then measured from an unweighted, least-squares fit
to the logarithmically binned data.
The PDS were not corrected for noise because the variability between
different orbits was much larger than the Poisson noise (as shown in the
previous section).
The 0.85~keV and 5~keV PDS for \ark\ and \tons\ are shown in Figures 6a
and 6b, respectively.

For \ark, the PDS changes monotonically from the softest (0.85~keV) band,
for which the slope of the PDS was $-1.22 \pm 0.06 $, and the hardest
(5~keV) band, for which the slope was $ -0.96 \pm 0.07 $.
Similar behavior was seen in \tons, which had a PDS slope of
$ -1.61 \pm 0.08 $ at 0.85~keV and $ -1.18 \pm 0.09 $ at 5~keV.
In both cases the slope differences are highly significant.
The sense of the difference is that the softest bands show more power on
the longest time scales probed.

\placefigure{fig6.ps}

\subsection{ Fractional Variability versus Flux Level }
\label{ Fvarflux }

For \ark, $\fv$s were also measured for each of the 256 orbits with more
than 32~min of data.
These were sorted by flux levels and averaged in flux bins with 20 or more
points in order to smooth out fluctuations.
The result is plotted as a function of mean count rate in Figure~7.
The variability amplitude is quite independent of count rate over a factor
of $\sim$4 in count rate, which means that the intrinsic RMS amplitude
(corrected for the measurement noise) is linearly correlated with flux.
The implications of this result are discussed in detail in \S~4.

\placefigure{fig7.ps}

\subsection {Linearity of the Light Curves }

A search for non-linear behavior (e.g., Leighly \& O'Brien 1997; Green,
McHardy \& Done 1999) was undertaken with the \ark\ and \tons\ light
curves.
The surrogate data method of Theiler \et (1992) was used, in which a
discriminating non-linear statistic is applied both to the real data and
to simulated light curves.
A significant difference between the values of the statistic as computed
for the real and simulated data indicates a detection of non-linearity in
the real light curve.
Here, the Kolmogorov-Sminov (KS) D-statistic, which compares the
distribution of data points above the mean with those below the mean
(e.g., Press \et 1992), is applied to all light curves.
A larger value of the D-statistic implies stronger non-linearity.

For each of the two targets, 100 simulated light curves, each with a PDS
slope corresponding to the PDS slope measured for the actual data, were
randomly generated using the algorithm of Timmer \& K\"{o}nig (1995).
Parent light curves had 4096 data points (much more than in the
observation to reduce red-noise leak) and a time resolution corresponding
to 1 \asca\ orbit.
A section of the light curve corresponding to the observation length was
randomly chosen and sampled in the same fashion as the actual data.
The KS D-statistic was calculated for each simulated light curve, and
these values were ranked.

The KS D-statistic for the summed 0.7--10 keV \ark\ light curve was found
to be greater than 86$\%$ of the KS D-statistic values for light curves
simulated with PDS slope of $-$1.13.
The KS D-statistic for the summed 0.7--10 keV \tons\ light curve was found
to be greater than 63$\%$ of the KS D-statistic values for light curves
simulated with PDS slope of $-$1.52.
Neither of these are $ > 1.5 \sigma $ effects.
Thus, this test provided no evidence for non-linear variability in either
of these light curves.
However, tests for non-linearity (and the related non-Gaussianity) are
notoriously difficult (see Press \& Rybiki 1997 for a detailed discussion)
so this is perhaps not as different from previous results as it might
appear.

\subsection{ Interband Lags }
\label{ lags }

Interband lags were searched for using the cross-correlation functions:
both the discrete correlation function (Edelson \& Krolik 1988) and
interpolated correlation function (White \& Peterson 1994).
The results are shown in Figure~8, and summarized in Tables~4 and 5.
They confirm that all of the \ark\ data are highly correlated, with
correlation coefficients $ r = 0.85 $ to 0.99, and none of the bands
appears to lead another, down to $\ls$1 orbit, or $ |\tau| <$~1.5~hr.
(The formal errors were much smaller, but we conservatively claim no limit
stronger than this; see Edelson \et 2001 for a detailed discussion of
uncertainties of interband lags and the perils of ``super-resolution".)
The \tons\ data are also highly correlated, although not nearly as well as
for \ark.
These data also show no lags down to limits of $\ls$1 orbit.
For CCFs that do not include the \euve\ data, correlation coefficients are
$ r = 0.48 $ to 0.92.
The \euve\ data is not as well correlated; for CCFs that do include the
\euve\ data, correlation coefficients were much lower: $ r = 0.32 $ to
0.59.

\placefigure{fig7.ps}

\placetable{table5}

\placetable{table6}

\section{ Discussion }
\label{ disco }

\subsection{ Separating Emission Components with Spectral Variability }

This monitoring of the soft Seyfert 1s \ark\ and \tons\ on time scales of
weeks revealed a number of new and interesting results:
on short time scales, the variations are similar in all bands, with
no measurable interband lags down to the shortest time scales measurable
and no consistent trend for the spectrum to harden or soften during flares
and dips.
However, especially in \ark, the hard and soft bands appear to diverge on
longer time scales, and the soft bands had slightly larger variability
amplitudes that apparently resulted from a long-term trend relative to the
hard bands.

It is difficult to see how a single emission component could naturally
produce spectral evolution that is so markedly different on long and short
time scales.
Instead, the simplest explanation is that two separate continuum emission
components are visible in the X-rays:
the first is a rapidly-variable hard component that dominates the
emission, especially at the hardest energies, for which the shape changes
only weakly, hardening slightly as the total flux changes by a factor of
$>$2.
The second is a much more slowly-variable ``soft excess" component only
seen in the lowest-energy channels of \asca.
Because it only contributes to the softest channels, these data alone
cannot determine if its shape changes with time.
There appears to be no obvious temporal connection between the two
components.

The spectral and variability properties of the soft component are not
consistent with the simplest models of direct thermal emission from an
optically thick, geometrically thin accretion disk (e.g., Frank, King \&
Raine 1992).
Even for the most favorable realistic parameters, the disk temperature is
well below 0.1~keV, while the observed emission (from the spectral fits)
extends well above 1~keV.
This general problem is well known (e.g., Czerny \& Elvis 1989).
While gravitational focussing and Comptoniation could harden the spectrum
somewhat, it is difficult to see how such a strong effect could be
produced.
Likewise, the relevant time scale for variations in a disk is probably the
viscous time scale, which for any reasonable set of parameters is years,
compared with the observed variability on time scales of $\ls$1 week (see
also Turner \et 2001b).

The hard component is generally identified with emission from a patchy
corona (e.g., Haardt, Maraschi \& Ghisellini 1994).
Because the cells are relatively small compared to the overall structure,
and the process could proceed as quickly as the light-crossing time (e.g.,
in the case of magnetic reconnection), then the expected time scales are
comfortably consistent with the observed variability time scales.

However, these results are not entirely consistent with the simple picture
in which the spectrally-defined fit parameters fully describe the
physically relevant emission components.
Turner \et (2001b) fitted the spectrum of \ark\ with a power-law and
(Gaussian) soft excess, and found that the soft excess component faded by
a factor of 2.8 throughout the observation, while the harder power-law
faded by only a factor of 1.68.
This is consistent with the overall hardness ratio changes reported in
\S~3.3.
However, this slowly-varying component would be nearly constant during a
single orbit, and thus would provide a constant ``contamination" at soft
energies during any rapid flares/dips.
This would yield a correlation between hardness and flux, in the sense
that the source would get harder during a flare and softer during a dip.
As discussed in \S~3.4., this is not the case.
This means that the straightforward spectral fits do not tell the full
story, and that most likely the rapidly variable component contains not
only the hard component (described as a power-law) but also some of the
soft excess as well.
That is, the hard component appears to be intrinsically more complex than
the pure power-law description used in spectral fitting routines.

\subsection{ Implications of Rapid Variability }

\ark\ shows factor of 2 flares and dips on time scales as short as
1000~sec.
For its redshift of $ z = 0.0247 $ (Huchra, Vogeley \& Geller 1999), this
corresponds to a change in the 0.7--10~keV luminosity $ \Delta L/\Delta t
\approx 10^{41} $~erg/s$^2$.
Under the assumptions of isotropic emission, the Eddington limit implies $
M_{BH} \ge 8 \times 10^5 L_{44} M_\odot $, where $L_{44}$ is the {\it
bolometric} luminosity in units of $10^{44}$~erg sec$^{-1}$ (e.g.,
Peterson 1997).
If we assume that \ark's 0.6--10~keV luminosity is 10\% of bolometric,
then $ L_{44} \approx 8 $ and $ M_{BH} \ge 6 \times 10^{6} M_\odot $.
We note that all of these assumptions mean that the limit is probably good
to no better than an order of magnitude.
Even so, for a black hole mass above this limit, both the radial
drift/viscous and thermal processes, operating at distances of $ \sim 10
R_S $, give time scales that are much too long (hours to years; see Frank
\et 1992) to be compatible with the observed time scale of $\sim$1000~sec.
Thus, such processes cannot be responsible for the observed X-ray
emission.

The light crossing time scale yields an approximate upper limit (to within
the order of magnitude uncertainties discussed above) on the size of the
emitting region of $ R \ls 15 R_{S} $ for $ M_{BH} \ge 6 \times 10^{6}
M_\odot $ and $ T = 1000 $~sec.
For other processes (governed, e.g., by the orbital or dynamical time
scales), the upper limit on the size of the emitting region must be
significantly smaller.
This indicates that the bulk of the X-ray emission in \ark\ must either be
produced in the inner accretion disk or else isolated clumps that are
smaller than or of order a few tens of Schwarzschild radii.

\subsection{ Statistical Properties of the X-ray Variability }

Decomposition of the X-ray emission into two components with very
different spectral shapes and variability time scales would also
significantly affect the interpretation of the PDS.
Recent intensive and long-term monitoring of Seyfert~1s have begun to
yield evidence that the power-law PDS measured at short time scales (e.g.,
Lawrence \& Papadakis 1993) show a turnover at longer time scales (Edelson
\& Nandra 2000, Pounds \et 2001, Uttley, McHardy \& Papadakis 2001).
However, the shape of this turnover is unclear, and it is consistent with
a variety of shapes (Uttley \et 2001).
The PDS of Galactic X-ray binaries (XRBs), for which the shapes are much
better-defined than for Seyfert~1s (due to their much shorter time scales
and higher fluxes) often show a more complex structure with multiple
features (Nowak 2001).
These multiple features could be multiple time scales, indicating that the
PDS cannot be modeled by single variability component.
The spectral evidence presented in this study suggests that the same
situation may be the case with Seyfert~1s.

The fact that $\fv$ is independent of flux level demonstrates that the
light curve is non-stationarity but in a relatively ``well-behaved" and
repeatable fashion.
This result confirms and expands upon the finding of Uttley \& McHardy
(2001) that found a similar independence of $\fv$ from flux for three
other Seyfert~1s, although those were measured with only two independent
flux points.
These results are consistent with no zero-point offset, indicating that
source does not have a large, constant flux component.
More importantly, the independence of $\fv$ from flux level shows yet
another remarkable parallel between Seyfert~1s and XRBs, suggesting a
relationship exists between these putative accreting black hole sources
independent of the mass of the central object, even though they differ by
a factor of $ \gs 10^6 $ in luminosity and black hole mass.

Finally, it is interesting that neither of these two objects show strong
non-linear variability, at least using the method of Theiler \et (1992).
Although this may not be the ideal method to use, visual examination of
the light curves also suggest that the variations are not wildly
non-linear, as the dips are about as strong as the flares (on a
logarithmic plot).

\subsection{ Soft and Hard Spectrum Seyfert 1 Galaxies }

With the study of these two soft Seyferts, it is now becoming feasible to
systematically explore the differences between the variability in soft and
hard Seyferts.
It is already well known that soft Seyferts tend to have narrower optical
permitted emission lines (Boller \et 1996) and that they show much
stronger X-ray variability than hard Seyferts at a similar luminosity
(Leighly 1999; Turner \et 1999a).
This may be more pronounced on short time scales: Pounds \et (2000) finds
that the PDS of \ark\ is unusually flat, meaning that there is more
variability power on short time scales relative to long time scales than
in hard Seyferts.

This is the first study to quantify the rapid variations in individual
sources: significant variations are almost always seen within a single
orbit ($\ls$40~min on source), and in the best studied case, \ark\ showed
factor of two variations in $\sim$6\% of all well-determined orbits.
This result is consistent with the idea first put forward by Pounds \et
(1995) that soft Seyferts are accreting at a much higher fraction of the
Eddington rate than hard Seyferts.

A new clue that is emerging involves the spectral variability results in
\S~3.1.: both of these soft Seyferts show only a very weak dependence of
variability amplitude on energy.
Other observations of soft Seyferts appear to show the same behavior.
A recent \chandra\ observation by Collinge \et (2001) found that the soft
Seyfert NGC~4051 varied in 0.5--8~keV flux by a factor of $>$5 in a
$\sim$4~ksec period while the 0.5--2~keV/2--8~keV flux ratio changed by
less than 20\%.
Likewise, Gliozzi \et (2001) find that the soft Seyfert PKS 0558--504
actually hardens as it gets brighter.
Finally, \xmm\ observations of both \tons\ and the soft Seyfert 1H
0707--495 show strong variability with almost no energy dependence
(Vaughan 2001).
This is very different behavior than seen in hard Seyferts, which
generally have X-ray spectra that appear to soften as they brighten (e.g.,
Markowitz \& Edelson 2001).

It is difficult to construct unified phenomenological picture that can
neatly explain all of these results.
Soft Seyferts tend to have stronger soft excesses, stronger overall
variability than hard Seyferts, yet hard Seyferts show much stronger
energy dependence of the variations (in the sense that their spectra
become softer as the flux increases).
This is the opposite of what would be expected from mixing a soft (rapidly
variable) and hard (less variable) component such that the former
dominates in soft Seyferts and the latter in hard Seyferts.
(It also contradicts the observation that the soft component appears to be
the less variable one.)
Likewise, if the harder component is the highly variable one, then one
would expect hard Seyferts to show stronger variability than soft
Seyferts.
Of course, it may be that these objects are may be powered by completely
different processes, and no unified scheme is applicable.

\section{ Summary }

This paper reports the most intensive X-ray monitoring ever undertaken of
any soft Seyfert galaxy.
These extraordinary data sets allow a deeper and more systematic
quantification of soft Seyfert variability than was previously possible.
Both sources show strong variability, with \ark\ showing repeated
variations of a factor of 2 on time scales as short as $\sim$1000~sec.
However, these relatively well-sampled light curves do not clear evidence
of non-linear behavior reported for other soft Seyferts, as the number and
strength of flares and dips were comparable.
The hard and soft light curves track well on short time scales, with no
clear trends for the hardness ratio to change in a systematic way during a
flare.
On longer time scales, especially for \ark, the hard and soft
bands diverge somewhat, yielding larger long time scale variability
amplitudes in the softer bands.

The rapid variations rule out thermal and viscous processes and constrain
the emission to the inner $\ls 15 R_S $, most likely to the inner disk or
small clumps in a corona.
The spectral variability indicates the presence of two components, the
dominant one (in the 0.6--10~keV \asca\ band) being a hard, rapidly
variable component that is naturally associated with a corona.
However, the softer, more slowly variable component cannot be identified
with the simplest optically thick, geometrically thin accretion disk
models, as the emission is observed to extend well beyond 1~keV, and the
observed variability time scales are much too short.

The variability amplitude was found to be almost independent of energy
band for these objects, and there are indications that the same is true
for other soft Seyferts as well.
This indicates a possibly important difference with hard Seyferts, which
generally show significantly softening of the spectrum as the flux
increases.
Other known differences between the sources is that soft Seyferts tend to
be more rapidly variable and also to have narrower optical permitted
lines.
This is not easy to understand in terms of phenomenological models in
which essentially identical hard and soft components are mixed together in
different ratios to produce the two types of Seyferts, and instead appear
to require a more complex explanation.

These observations show that intensive spectral variability monitoring has
unique power to separate out emission components in a way that is
complementary to single-epoch spectroscopy.
As these objects are much too distant to image directly, spectral
variability studies may prove our most effective tool for determining the
processes responsible for the high X-ray luminosities of AGN.
While the current \asca\ and \xte\ archives contain a great deal of
relevant data, we expect that future progress will hinge on \xmm.
Its high throughput makes it the only instrument with sufficient
sensitivity to obtain meaningful short term light curves for the most
extreme and interesting soft Seyferts which tend to be almost an order of
magnitude fainter than \ark\ and \tons.
Its broad bandpass allows it to simultaneously study spectral variations
over a much larger fraction of the X-ray spectrum than was previously
possible, especially at the critical soft energies (which were not probed
by \asca\ or \xte).
Finally, its high-Earth orbit yields uninterrupted $\sim$40~hr light
curves that can be used to study short time scale variability.
The previous generation of low-Earth orbit telescopes produced light
curves corrupted by interruptions that made it impossible to track the
development of flares.
Although the ideal parameters for such observations are not yet fully
determined, it is likely that more insight will be gained from a few long
observations instead of many short ones (Mushotzky 2001).
It is also important that future variability studies accurately define the
variability properties of both soft Seyferts (especially the most extreme
examples like IRAS~13224--3908, PHL~1092 and 1H~0707-495) but also of a
control of group of ``standard" hard Seyferts such as NGC~5548.
As such long observations are unlikely to be scheduled in great numbers in
this early stage of the mission (e.g., only one Seyfert 1, MCG--6-30-15,
has been scheduled for more than a single orbit in the first two years of
\xmm), patience is a necessary virtue in this area of endeavor.

\acknowledgements
The authors thank the \xte\ and \asca\ teams for their efforts that
resulted in the data needed for this research.
They also thank the referee, Niel Brandt, for helping to focus the
discussion on the big picture.
Edelson and Markowitz were supported by NASA grants NAG~5-7317 and
NAG~5-9023, and Turner was supported by NASA grant NAG~5-7385.

\newpage

\appendix

\section{ The Estimation of the Fractional Excess Variability Amplitude,
$\fv$, of an AGN Light Curve }

Here we present a prescription for measuring the fractional excess
variability parameter $\fv$ and its associated error. We also note
various caveats relating to its interpretation.

\subsection{ Basic Equations and Derivation of $\fv$ }

Consider a light curve subdivided into $N$ time bins, where each bin is
further subdivided into $n_i$ individual points ($n_i$ can be the same or
different in each bin). The mean count rate in the $ith$ bin is:
\begin{equation}
X_i = { 1 \over n_i } \sum_{j=1}^{n_i} x_{ij},
\end{equation}
where $x_{ij}$ is the count rate of the $jth$ point in the $ith$ bin.
The square of the standard error on $X_i$ is:
\begin{equation}
\sigma^2_{err,i} = { 1 \over n_i(n_i-1) } \sum_{j=1}^{n_i}
(x_{ij}-X_i)^2 .
\end{equation}

In considering the full light curve, the unweighted mean count rate
given by:
\begin{equation}
\langle X \rangle = { 1 \over N } \sum_{i=1}^N X_i ,
\end{equation}
and the variance of the binned data comprising the light curve is:
\begin{equation}
S^2 = { 1 \over N-1 } \sum_{i=1}^N (X_i-\langle X \rangle)^2 .
\end{equation}

Both intrinsic source variability and measurement uncertainty contribute
to this observed variance.
Under the assumption that both components are normally distributed and
combine in quadrature, the observed variance can be written as:
\begin{equation}
S^2 = \langle X \rangle^2 \fxv + \langle \sigma_{err}^2 \rangle
\end{equation}
The first term on the right represents the intrinsic scatter induced
by the source variability. The second term is the contribution of the
measurement noise.  We assume that the scatter of the data points
within an individual time bin is predominantly due to the statistical
uncertainty of the measurements, leading to:
\begin{equation}
\langle \sigma_{err}^2 \rangle = { 1 \over N }
\sum_{i=1}^N\sigma^2_{err,i},
\end{equation}
Rearranging equation A5 yields the standard definition for the fractional
excess variance
\begin{equation}
\fxv = { S^2 - \langle \sigma_{err}^2 \rangle \over \langle X \rangle^2 }.
\end{equation}
The fractional variability amplitude $\fv$ is simply the square root of
the
fractional excess variance:
\begin{equation}
\fv = \sqrt{ S^2 - \langle \sigma_{err}^2 \rangle \over \langle X
\rangle^2 },
\end{equation}
as given in Equation~1 of the text.

\subsection{ Derivation of the Uncertainty on $\fv$ }

We now require a measure of the uncertainties that should be assigned
to $ \xsv $ and $ \fv $.
In equation A7, assume that the dominant variance will be that associated
with the quantity $S^2$, and that the error term $\langle \sigma_{err}^2
\rangle$ can be neglected by comparison.
The implications of this assumption are discussed at the end of this
section.

This variance on $S^2$ can be estimated as $ {2 \over N-1} S^4 \approx {2
\over N} S^4 $ (e.g., Trumpler \& Weaver 1962).
Hence the standard deviation of $\fxv$ is:
\begin{equation}
\sigma_{\fxv} = \sqrt{2 \over N} {S^2 \over \langle X \rangle^2}
\end{equation}
Setting $ x = \xsv $ and $ y = \fv $ so that $ y = \sqrt{x} $ yields
\begin{equation}
{ dy \over dx } = { 1 \over 2 \sqrt{x} } = { 1 \over 2 y }
= { 1 \over 2 \fv }
\end{equation}
Transmitting the error through the equation by the standard formula
$ \sigma_y = { dy \over dx } \sigma_x $ yields
\begin{equation}
\sigma_{\fv} = { 1 \over 2 \fv } \sigma_{\xsv}
= {1 \over 2 \fv} \sqrt{1 \over N} {S^2 \over \langle X \rangle^2}
\end{equation}
as in Equation~2 of the text.

In the above analysis the assumption (made in eqn. A2) that all of the
variance within a time bin is due solely to measurement errors will lead
to overestimation of the latter if the source exhibits rapid variability
on time scales comparable to the bin size.
This is a conservative approach which in many circumstances may be a
better choice than relying on the errors propagated through data
extraction and data
fitting algorithms (which may mix systematic and statistical errors in a
manner not appropriate for variability studies).
The importance of such an approach can be seen in the fact that the error
estimate assumed that the variance due to systematic errors was small
compared to the total variance; if they are not the derivation is
incorrect.

More serious, however, is the assumption that the underlying source
variability is governed by processes that are stationary and governed by
Gaussian statistics.
As red-noise processes are "weakly non-stationary" (e.g., Press \& Rybicki
1997) the above error estimate cannot account for random fluctuations in
$\fv$ as a function of time.
A further point is that the weak non-stationarity and (in general)
non-normal distribution of fluxes in red-noise light curves mean that the
above prescription provides an increasingly poor estimate of the
uncertainty on $\fv$ as the signal-to-noise in the observed light curve
increases.
(This will be discussed in more detail in a future work, Vaughan \et in
prep.)
A more robust approach would be to estimate the PDS, but where this is not
possible $\fv$ can provide a useful measure of the degree of variability
in a given light curve.
In practice the value of statistics such as $F_{var}$ is as a comparative
measure of the magnitude and constancy of the variability signal.

\begin{deluxetable}{llcccc}
\tablewidth{5.9in}
\tablenum{1}
\tablecaption{ Observing Log \label{tab1}}
\small
\tablehead{
\colhead{ } & \colhead{ } & \colhead{ Energy } &
\colhead{ JD Range } & \colhead{ Sampling } & \colhead{ Number } \\
\colhead{ Source } & \colhead{ Instrument } & \colhead{ Range (keV) } &
\colhead{ (--2,450,000.5) } & \colhead{ Rate (min) } & \colhead{ Points }
}
\startdata
\tons & \asca\ SIS &  0.7 -- 10 & 1515.55 -- 1527.78 & 94.8  & 180 \\
      & \asca\ GIS & 0.95 -- 10 & 1515.55 -- 1527.78 & 94.8  & 181 \\
      & \euve\     & 0.1 -- 0.2 & 1515.56 -- 1527.79 & 94.1  & 142 \\
      & \xte\      &   2 -- 10  & 1515.59 -- 1527.72 & 95.8  & 142 \\
\ark  & \asca\ SIS &  0.7 -- 10 & 1696.52 -- 1731.00 & 94.0  & 518 \\
      & \asca\ GIS & 0.95 -- 10 & 1696.52 -- 1731.00 & 94.0  & 520 \\
      & \xte\      &   2 -- 10  & 1696.63 -- 1726.47 & 191.4 & 231 \\
\enddata
\end{deluxetable}

\begin{deluxetable}{lccccccc}
\tablewidth{6.3in}
\tablenum{2}
\tablecaption{ \ark\ Variability Parameters \label{table2}}
\small
\tablehead{
\colhead{ } & \colhead{ Energy } & \colhead{ Band } & \colhead{ Count } &
\colhead{  } &
\colhead{ } & \colhead{  } & \colhead{ Point-to-} \\
\colhead{ Instru- } & \colhead{ Range} &
\colhead{ Center} & \colhead{ Rate  } & \colhead{ Signal } & \colhead{
Total } &
\colhead{ Fractional } & \colhead{ Point } \\
\colhead{ ment} & \colhead{ (keV) } & \colhead{ (keV) }
& \colhead{ (c/s) } & \colhead{ to Noise } & \colhead{ Variance }
& \colhead{ Variability } & \colhead{Variability } }
\startdata
\asca\ & 0.7 -- 0.95 & 0.85& 0.57 & 31.1 & 34.2 $\pm$ 1.1\%  & 34.0 $\pm$
1.1\%  & 14.4 $\pm$ 0.5\% \\
\asca\ & 0.95 -- 1.3 & 1.1 & 1.42 & 43.4 & 33.2 $\pm$ 1.0\%  & 33.1 $\pm$
1.0\%  & 15.5 $\pm$ 0.5\% \\
\asca\ & 1.3 -- 2    & 1.5 & 1.57 & 43.3 & 33.4 $\pm$ 1.0\%  & 33.3 $\pm$
1.0\%  & 15.9 $\pm$ 0.5\% \\
\asca\ & 2 -- 4      & 2.5 & 0.86 & 33.3 & 32.4 $\pm$ 1.0\%  & 32.3 $\pm$
1.0\%  & 16.0 $\pm$ 0.5\% \\
\asca\ & 4 -- 10     &   5 & 0.27 & 19.7 & 30.3 $\pm$ 0.9\%  & 29.8 $\pm$
1.0\%  & 16.8 $\pm$ 0.6\% \\
& & & & & & \\
SIS  & 0.7 -- 0.95 & 0.85 & 0.57 & 31.1  & 34.2 $\pm$ 1.1\%  & 34.0 $\pm$
1.1\%  & 14.4 $\pm$ 0.5\% \\
SIS  & 0.95 -- 1.3 & 1.1 & 1.00 & 39.7   & 33.2 $\pm$ 1.0\%  & 33.1 $\pm$
1.0\%  & 15.0 $\pm$ 0.5\% \\
SIS  & 1.3 -- 2    & 1.4 & 0.95 & 37.5   & 33.3 $\pm$ 1.0\%  & 33.2 $\pm$
1.0\%  & 15.6 $\pm$ 0.5\% \\
SIS  & 2 -- 4      & 2.5 & 0.45 & 27.0   & 32.6 $\pm$ 1.0\%  & 32.4 $\pm$
1.0\%  & 15.8 $\pm$ 0.5\% \\
SIS  & 4 -- 10     &   5 & 0.13 & 14.8   & 31.2 $\pm$ 1.0\%  & 30.4 $\pm$
1.0\%  & 16.4 $\pm$ 0.6\% \\
& & & & & & \\
GIS  & 0.95 -- 1.3& 1.1 & 0.42 & 28.6    & 33.0 $\pm$ 1.0\%  & 32.8 $\pm$
1.0\%  & 15.1 $\pm$ 0.5\% \\
GIS  & 1.3 -- 2   & 1.5 & 0.62 & 33.8    & 33.0 $\pm$ 1.0\%  & 32.8 $\pm$
1.0\%  & 15.4 $\pm$ 0.5\% \\
GIS  & 2 -- 4     & 2.5 & 0.41 & 27.5    & 32.0 $\pm$ 1.0\%  & 31.7 $\pm$
1.0\%  & 15.4 $\pm$ 0.5\% \\
GIS  & 4 -- 10    &   5 & 0.14 & 16.2    & 29.7 $\pm$ 0.9\%  & 29.0 $\pm$
0.9\%  & 16.3 $\pm$ 0.6\% \\
& & & & & & \\
\xte\ & 2 -- 4    & 3.3 & 0.78 & 14.0    & 34.6 $\pm$ 1.6\%  & 33.8 $\pm$
1.6\%  & 24.2 $\pm$ 1.2\% \\
\xte\ & 4 -- 10   &   6 & 1.04 & 13.4    & 33.5 $\pm$ 1.6\%  & 32.6 $\pm$
1.6\%  & 23.9 $\pm$ 1.2\% \\
\enddata
\end{deluxetable}

\begin{deluxetable}{lccccccc}
\tablewidth{6.3in}
\tablenum{3}
\tablecaption{ \tons\ Variability Parameters \label{table3}}
\small
\tablehead{
\colhead{ } & \colhead{ Energy } & \colhead{ Band } & \colhead{ Count } &
\colhead{  } &
\colhead{ } & \colhead{  } & \colhead{ Point-to-} \\
\colhead{ Instru- } & \colhead{ Range} &
\colhead{ Center} & \colhead{ Rate  } & \colhead{ Signal } & \colhead{
Total } &
\colhead{ Fractional } & \colhead{ Point } \\
\colhead{ ment} & \colhead{ (keV) } & \colhead{ (keV) }
& \colhead{ (c/s) } & \colhead{ to Noise } & \colhead{ Variance }
& \colhead{ Variability } & \colhead{Variability } }
\startdata
\euve\ & 0.1 -- 0.2 & 0.15 & 0.14 & 5.5   & 28.3 $\pm$ 1.7\%  & 17.8 $\pm$
2.7\%  & Undefined \\
& & & & & & \\
\asca\ & 0.7 -- 0.95 & 0.85 & 0.19 & 17.2 & 20.3 $\pm$ 1.1\%  & 19.4 $\pm$
1.1\%  & 7.2 $\pm$ 0.6\% \\
\asca\ & 0.95 -- 1.3 & 1.1 & 0.37 &  21.5 & 19.6 $\pm$ 1.0\%  & 19.0 $\pm$
1.0\%  & 9.4 $\pm$ 0.6\% \\
\asca\ & 1.3 -- 2    & 1.5 & 0.42 &  23.4 & 17.8 $\pm$ 0.9\%  & 17.2 $\pm$
1.0\%  & 9.6 $\pm$ 0.6\% \\
\asca\ & 2 -- 4      & 2.5 & 0.25 &  18.0 & 18.5 $\pm$ 1.0\%  & 17.5 $\pm$
1.0\%  & 10.4 $\pm$ 0.7\% \\
\asca\ & 4 -- 10     & 5   & 0.08 &  9.8  & 19.4 $\pm$ 1.0\%  & 16.2 $\pm$
1.2\%  & 9.3 $\pm$ 1.1\% \\
& & & & & & \\
SIS  & 0.7 -- 0.95 & 0.85 & 0.19 & 17.2   & 20.3 $\pm$ 1.1\%  & 19.4 $\pm$
1.1\%  & 7.2 $\pm$ 0.6\% \\
SIS  & 0.95 -- 1.3 & 1.1 & 0.28 &  20.8   & 18.6 $\pm$ 1.0\%  & 17.9 $\pm$
1.0\%  & 8.5 $\pm$ 0.6\% \\
SIS  & 1.3 -- 2    & 1.4 & 0.24 &  19.4   & 18.3 $\pm$ 1.0\%  & 17.5 $\pm$
1.0\%  & 9.8  $\pm$ 0.7\% \\
SIS  & 2 -- 4      & 2.5 & 0.13 &  13.9   & 19.1 $\pm$ 1.0\%  & 17.5 $\pm$
1.1\%  & 10.9 $\pm$ 0.8\% \\
SIS  & 4 -- 10     & 5   & 0.04 &  7.1    & 21.2 $\pm$ 1.1\%  & 15.5 $\pm$
1.5\%  & 8.6 $\pm$ 1.7\% \\
& & & & & & \\
GIS  &0.95 -- 1.3 & 1.1 & 0.09 &  11.4    & 24.7 $\pm$ 1.3\%  & 22.9 $\pm$
1.4\%  & 11.6 $\pm$ 1.0\% \\
GIS  & 1.3 -- 2   & 1.5 & 0.17 &  17.3    & 17.9 $\pm$ 0.9\%  & 16.8 $\pm$
1.0\%  & 9.4 $\pm$ 0.7\% \\
GIS  & 2 -- 4     & 2.5 & 0.12 &  14.6    & 18.7 $\pm$ 1.0\%  & 17.3 $\pm$
1.1\%  & 9.3 $\pm$ 0.8\% \\
GIS  & 4 -- 10    & 5   & 0.05 &  8.4     & 20.0 $\pm$ 1.1\%  & 15.8 $\pm$
1.3\%  & 9.1 $\pm$ 1.3\% \\
& & & & & & \\
\xte\ & 2 -- 4     & 3.3 & 0.11 &  6.1    & 26.2 $\pm$ 1.6\%  & 19.6 $\pm$
2.1\%  & 10.5 $\pm$ 2.3\% \\
\xte\ & 4 -- 10    & 6   & 0.16 &  5.8    & 28.9 $\pm$ 1.7\%  & 22.1 $\pm$
2.2\%  & 12.8 $\pm$ 2.3\% \\
\enddata
\end{deluxetable}

\begin{deluxetable}{cccccc}
\tablewidth{3.6in}
\tablenum{4}
\tablecaption{ \ark\ Cross Correlation Results \label{table4}}
\small
\tablehead{
\colhead{ Band 1 } & \colhead{ Band 2 } & \colhead{ DCF } &
\colhead{ DCF } & \colhead{ ICF } & \colhead{ ICF } \\
\colhead{ (keV) } & \colhead{ (keV) } & \colhead{ r$_{max}$ } &
\colhead{ $\tau$ (hr)  } & \colhead{ r$_{max}$ } & \colhead{ $\tau$ (hr) }
}
\startdata
0.85   &  1.1  & 0.97 & 0.0     & 0.97  & 0.0 \nl
0.85   &  1.5  & 0.94 & 0.0     & 0.94  & 0.0 \nl
0.85   &  2.5  & 0.91 & 0.0     & 0.92  & 0.0 \nl
0.85   &  5    & 0.85 & 0.0     & 0.85  & 0.0 \nl
1.1    &  1.5  & 0.99 & 0.0     & 0.99  & 0.0 \nl
1.1    &  2.5  & 0.97 & 0.0     & 0.97  & 0.0 \nl
1.1    &  5    & 0.91 & 0.0     & 0.91  & 0.0 \nl
1.5    &  2.5  & 0.99 & 0.0     & 0.99  & 0.0 \nl
1.5    &  5    & 0.94 & 0.0     & 0.94  & 0.0 \nl
2.5    &  5    & 0.97 & 0.0     & 0.97  & 0.0 \nl
\enddata
\end{deluxetable}

\begin{deluxetable}{cccccc}
\tablewidth{3.6in}
\tablenum{5}
\tablecaption{ \tons\ Cross Correlation Results \label{table5}}
\small
\tablehead{
\colhead{ Band 1 } & \colhead{ Band 2 } & \colhead{ DCF } &
\colhead{ DCF } & \colhead{ ICF } & \colhead{ ICF } \\
\colhead{ (keV) } & \colhead{ (keV) } & \colhead{ r$_{max}$ } &
\colhead{ $\tau$ (hr)  } & \colhead{ r$_{max}$ } & \colhead{ $\tau$ (hr) }
}
\startdata
0.15    &  0.85 & 0.59 & $-$1.6   & 0.63  &  0.0 \nl
0.15    &  1.1  & 0.54 & 0.0      & 0.61  & +0.8 \nl
0.15    &  1.5  & 0.46 & +1.6     & 0.54  & +0.8  \nl
0.15    &  2.5  & 0.44 & +1.6     & 0.49  & +1.6  \nl
0.15    &  5    & 0.32 & +4.8     & 0.35  & +4.0 \nl
0.85    &  1.1  & 0.92 & 0.0      & 0.92  & 0.0     \nl
0.85    &  1.5  & 0.83 & 0.0      & 0.83  & 0.0    \nl
0.85    &  2.5  & 0.77 & 0.0      & 0.77  & 0.0     \nl
0.85    &  5    & 0.48 & 0.0      & 0.49  & 0.0  \nl
1.1    &  1.5  & 0.92 & 0.0      & 0.92  & 0.0     \nl
1.1    &  2.5  & 0.84 & 0.0      & 0.84  & 0.0   \nl
1.1    &  5    & 0.61 & 0.0      & 0.62  & +0.8   \nl
1.5    &  2.5  & 0.90 & 0.0      & 0.90  & 0.0    \nl
1.5    &  5    & 0.72 & 0.0      & 0.72  & 0.0    \nl
2.5    &  5    & 0.72 & 0.0      & 0.73  & +0.8  \nl
\enddata
\end{deluxetable}

\newpage

\begin{figure}[ht]
\includegraphics{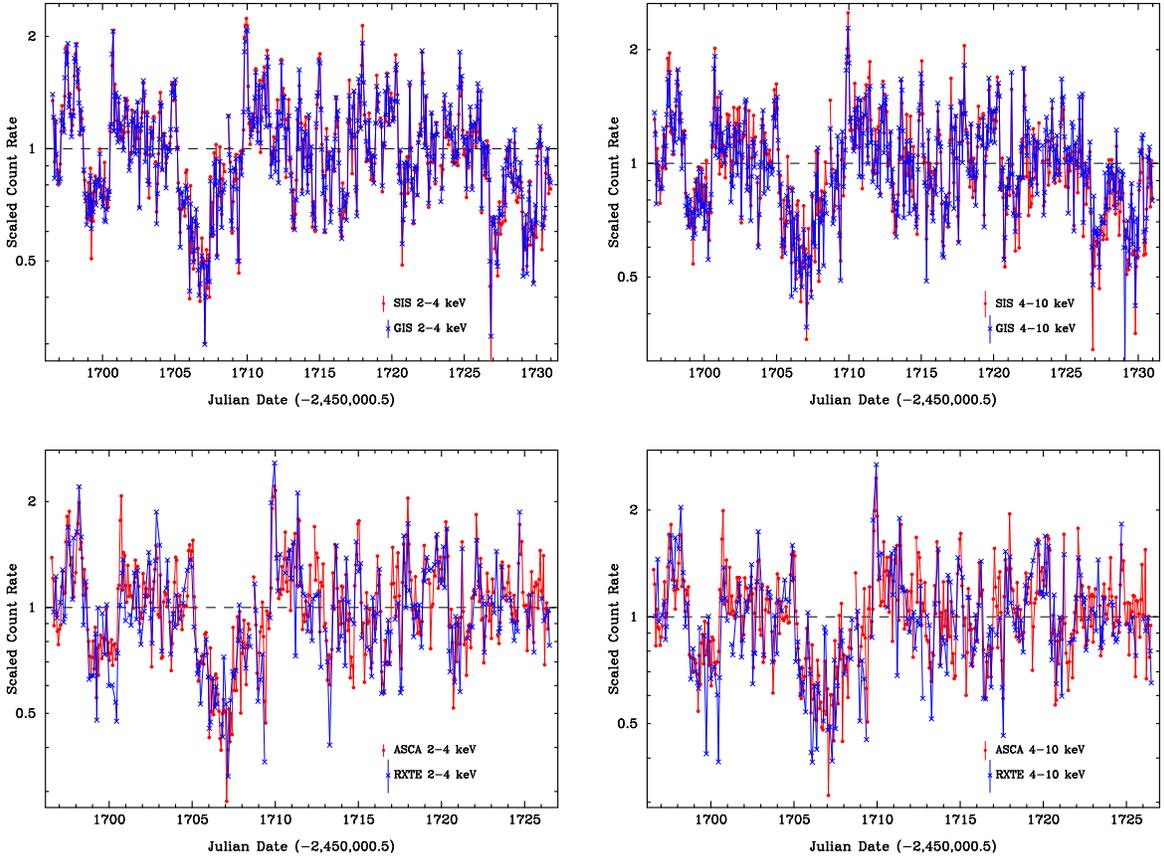}
\vspace{5.5in}
\caption{
Light curve overplot diagrams for \ark.
All data were scaled by dividing by the mean of that light curve to
eliminate the effects of differing instrumental sensitivities.
In the upper left, the 2-4~keV \asca\ SIS (red, circles) and GIS (blue,
triangles) light curves are shown,
in the upper right, the 4-10~keV SIS and GIS light curves,
in the lower left, the 2-4~keV summed \asca\ (red, circles) and \xte\
(blue, triangles) light curves,
and in the lower right, the 4-10~keV summed \asca\ and \xte\ light curves.
Note the good agreement between the \asca\ SIS and GIS data (also seen in
lower energy bands), while the \xte\ data does not agree as well.
}\label{fig1a}
\end{figure}

\setcounter{figure}{0}
\begin{figure}[ht]
\includegraphics{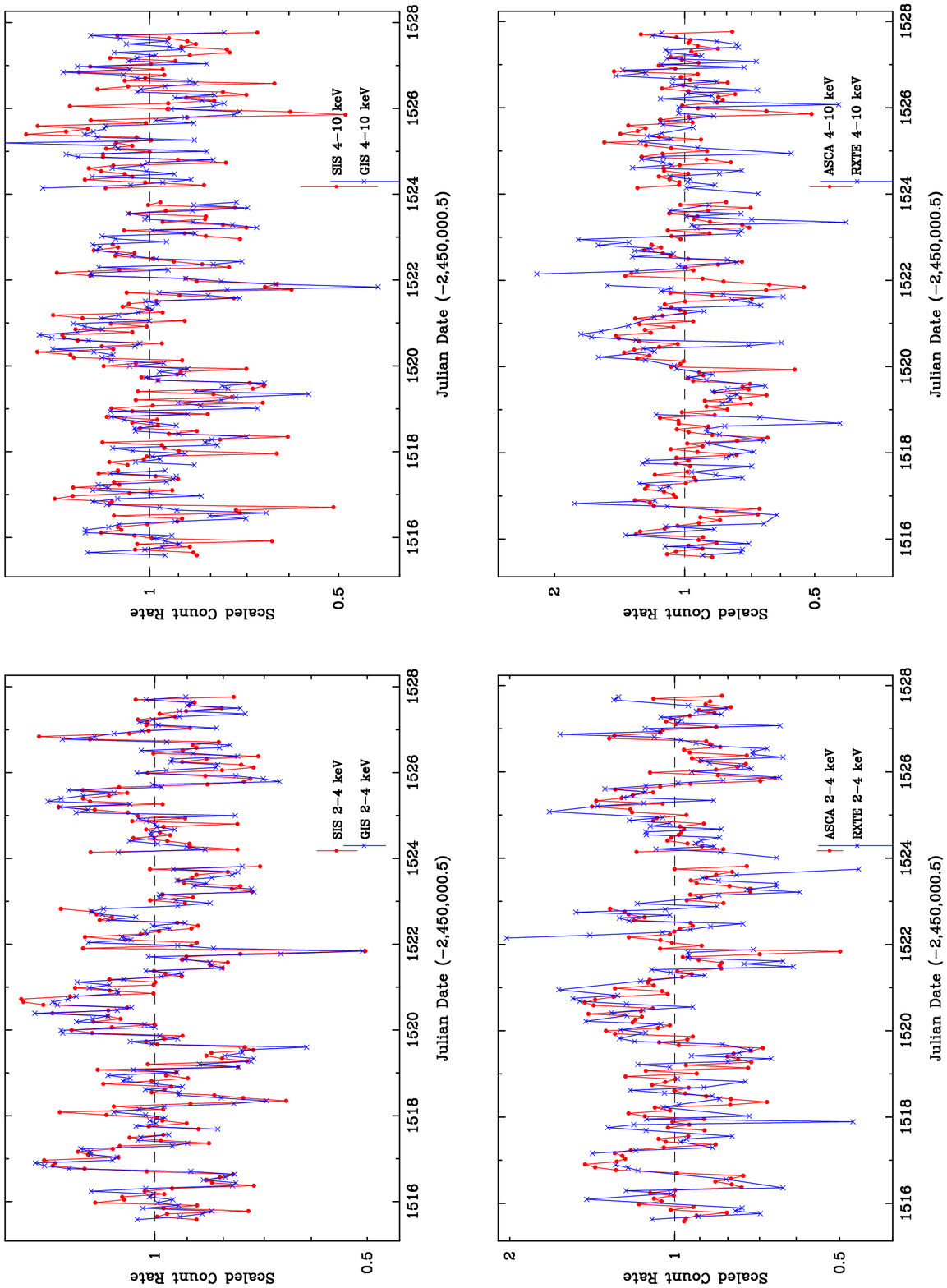}
\vspace{5.5in}
\caption{
Same as Figure~1a, but for \tons.
Note the good agreement between the \asca\ SIS and GIS data, but
relatively poor agreement between the \xte\ and \asca\ quasi-simultaneous
data.
}\label{fig1b}
\end{figure}

\begin{figure}[ht]
\includegraphics{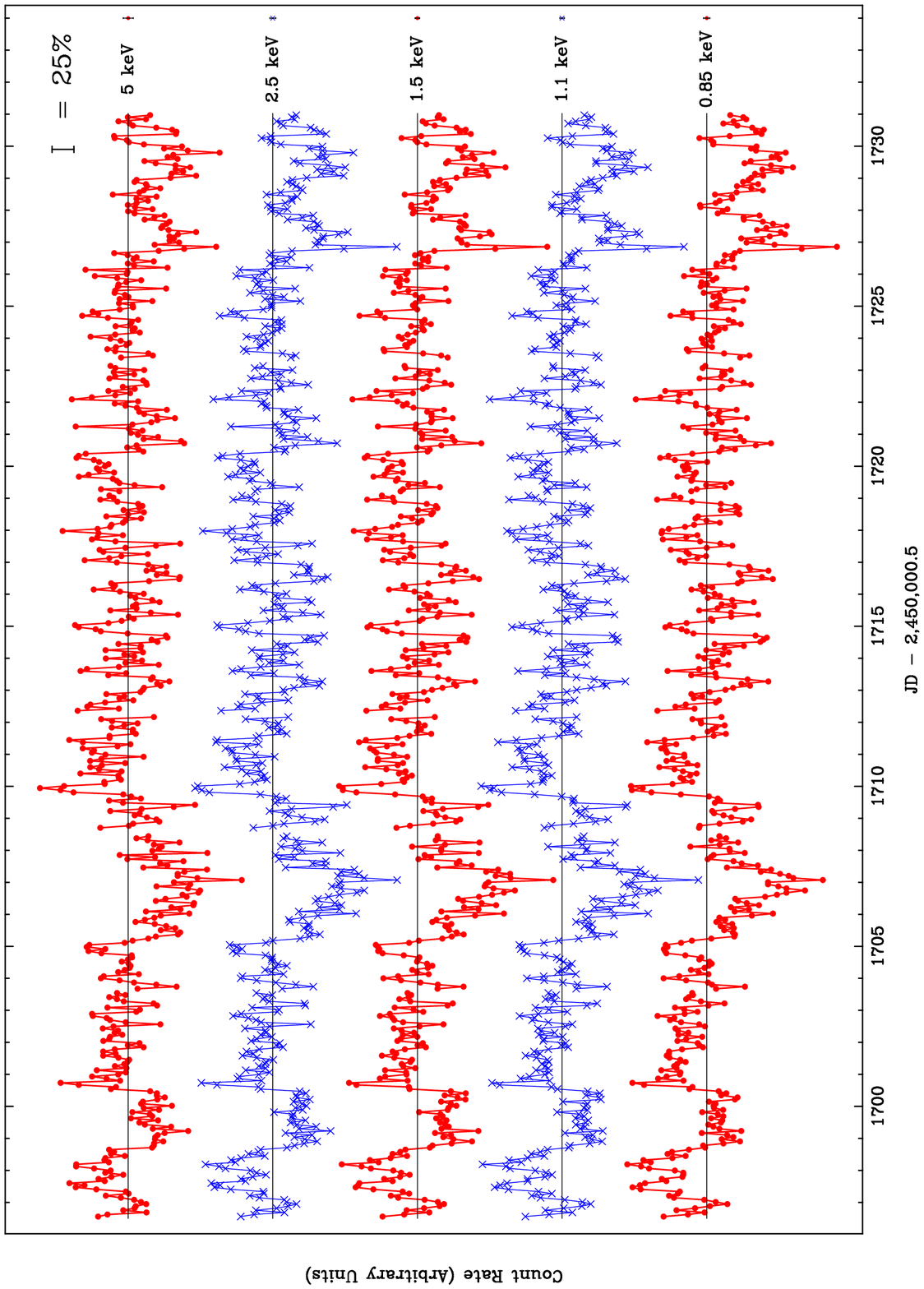}
\vspace{5.5in}
\caption{
Light curves for \ark, covering from top to bottom, 5~keV, 2.5~keV,
1.5~keV, 1.1~keV and 0.85~keV.
As the data are presented in logarithmic units, with an arbitrary offset
between bands, a 25\% change is shown in the upper right.
Error bars are not shown because the figure would become too crowded, so
typical 1$\sigma$ errors are shown on the right.
Lines connect the points only for adjacent orbits, so a broken line
indicates
that an orbit was missing.
}\label{fig2a}
\end{figure}

\setcounter{figure}{1}
\begin{figure}[ht]
\includegraphics{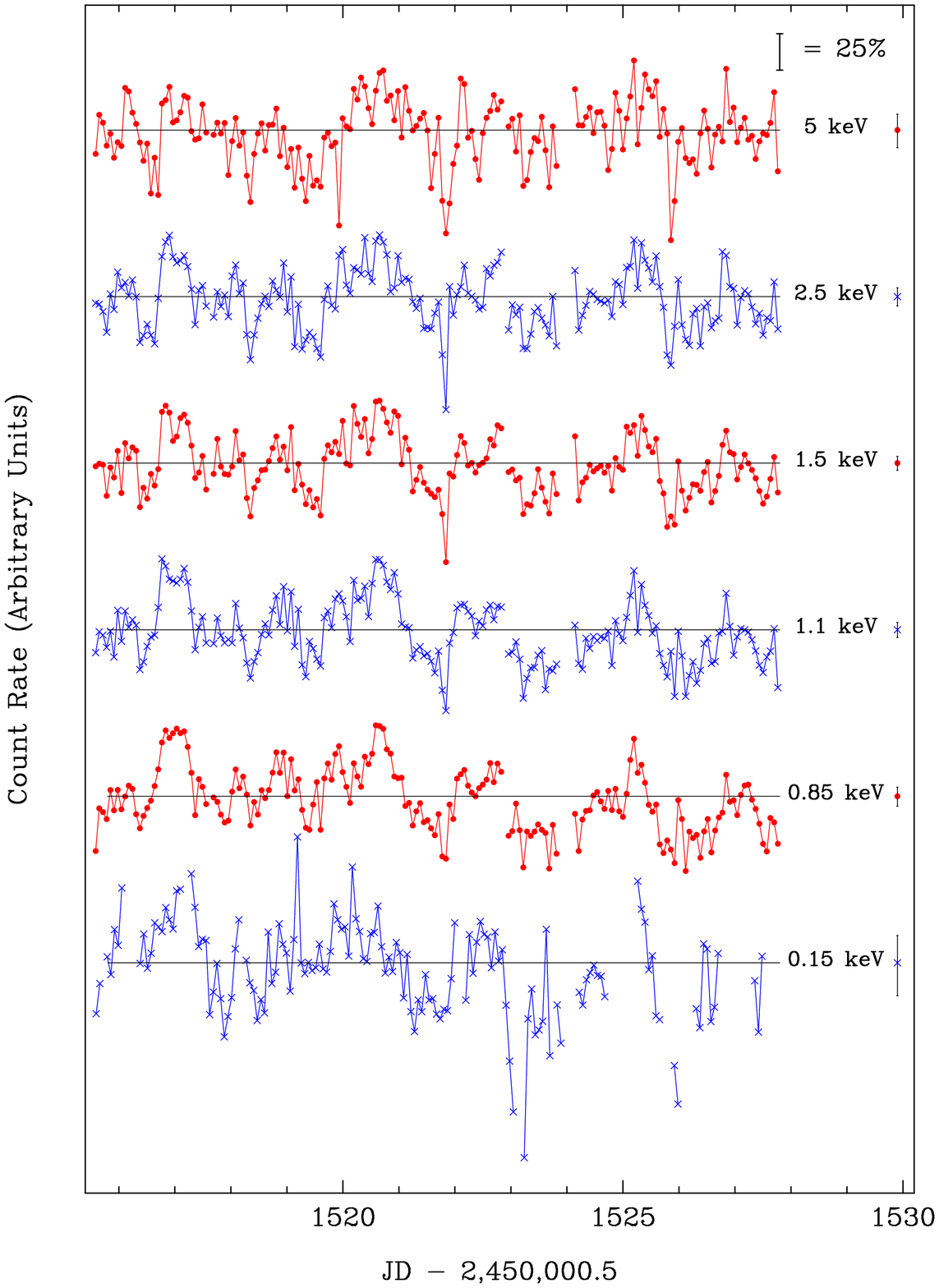}
\vspace{5.5in}
\caption{
Same as Figure~2a, but for \tons.
The bottom light curve is for \euve\ 0.15~keV.
}\label{fig2b}
\end{figure}

\begin{figure}[ht]
\epsscale{0.85}
\plotone{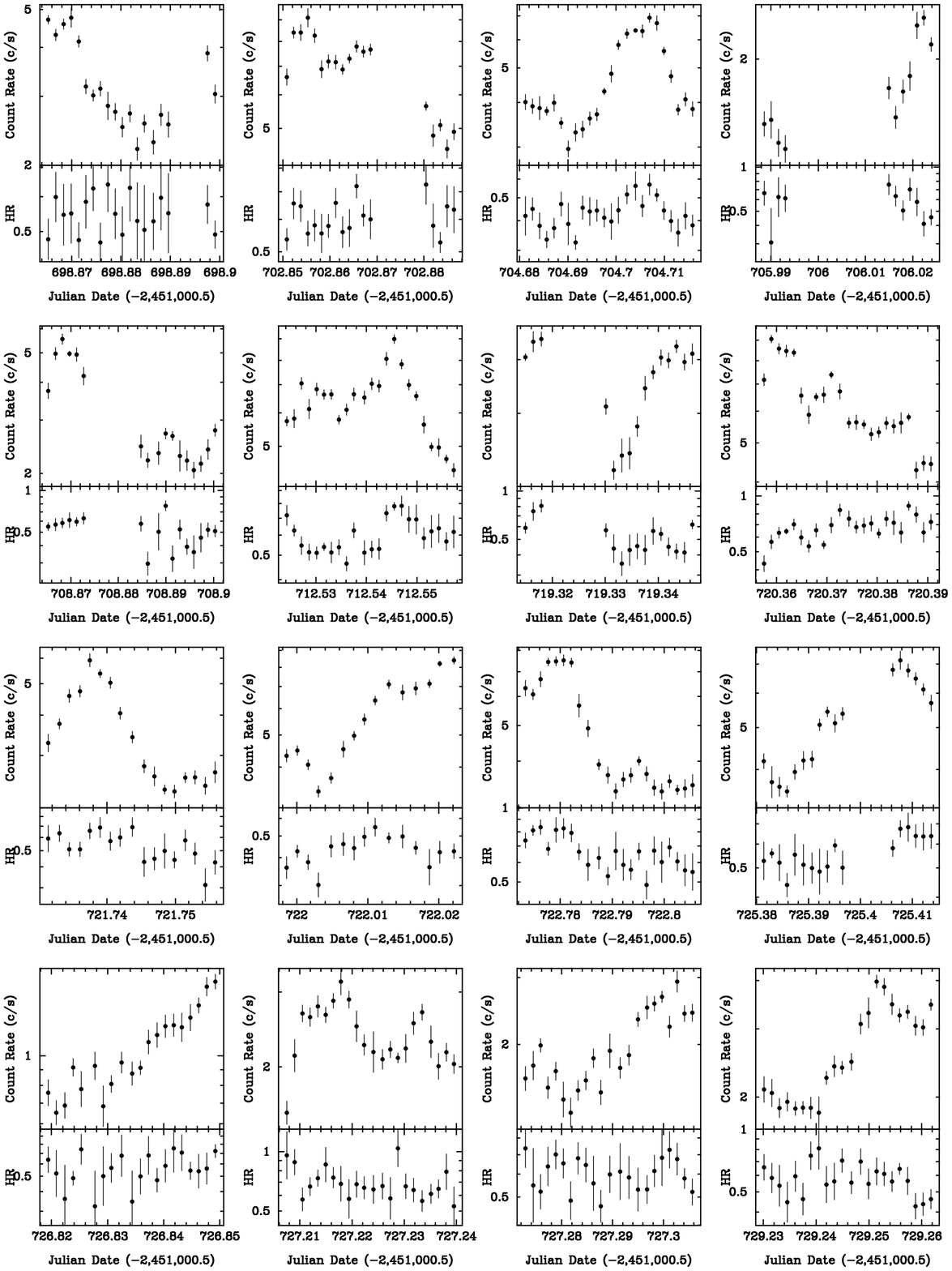}
\caption{
Single orbit light curves for \ark.
The total 0.7--10~keV count rate is shown on the top and the
2--10~keV/0.7--1.3~keV hardness ratio on the bottom.
The 16 orbits in which a factor of $\ge$2 flux variation are plotted.
The plots are ordered by time.
}\label{fig3}
\end{figure}

\begin{figure}
\epsscale{0.85}
\plottwo{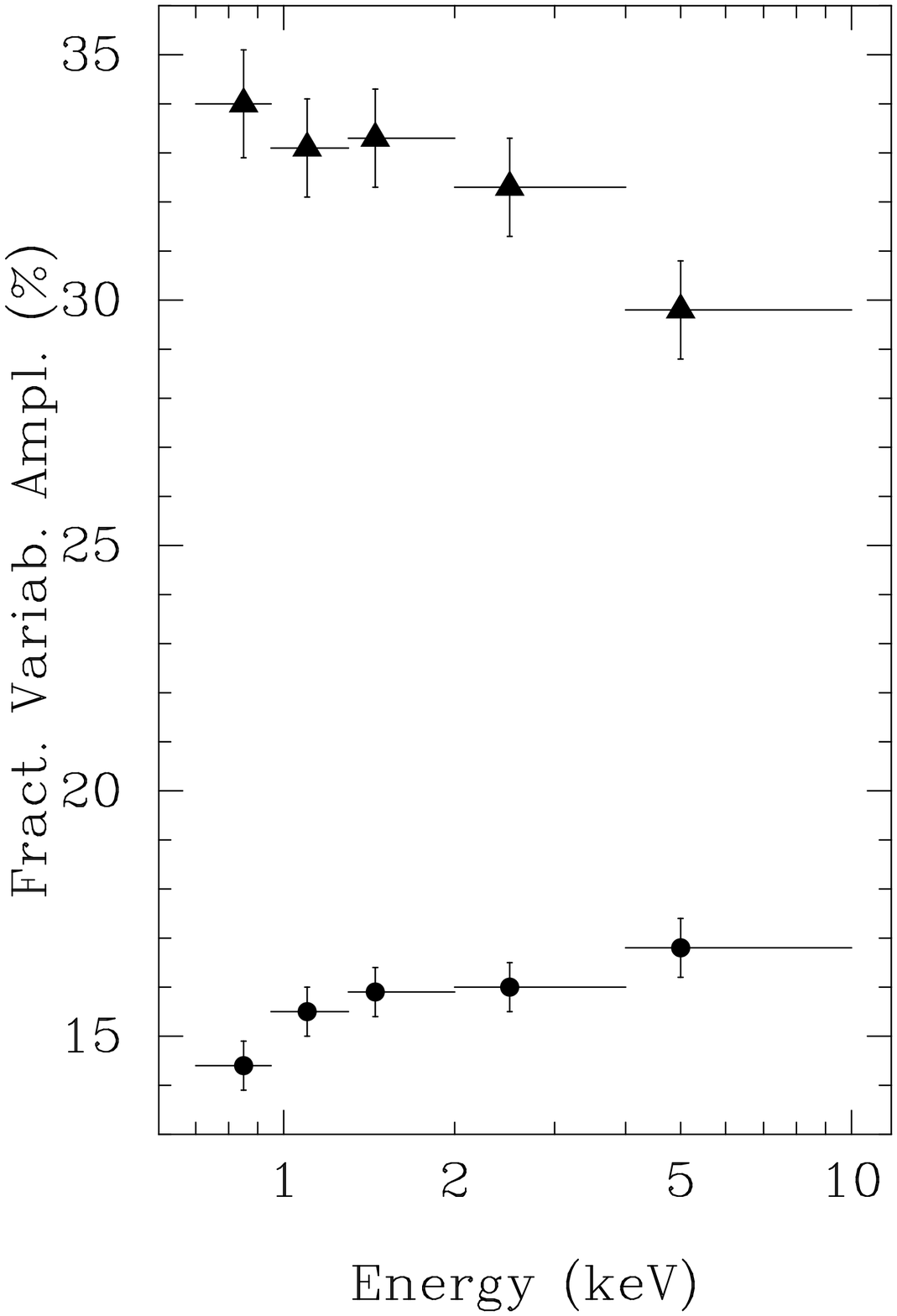}{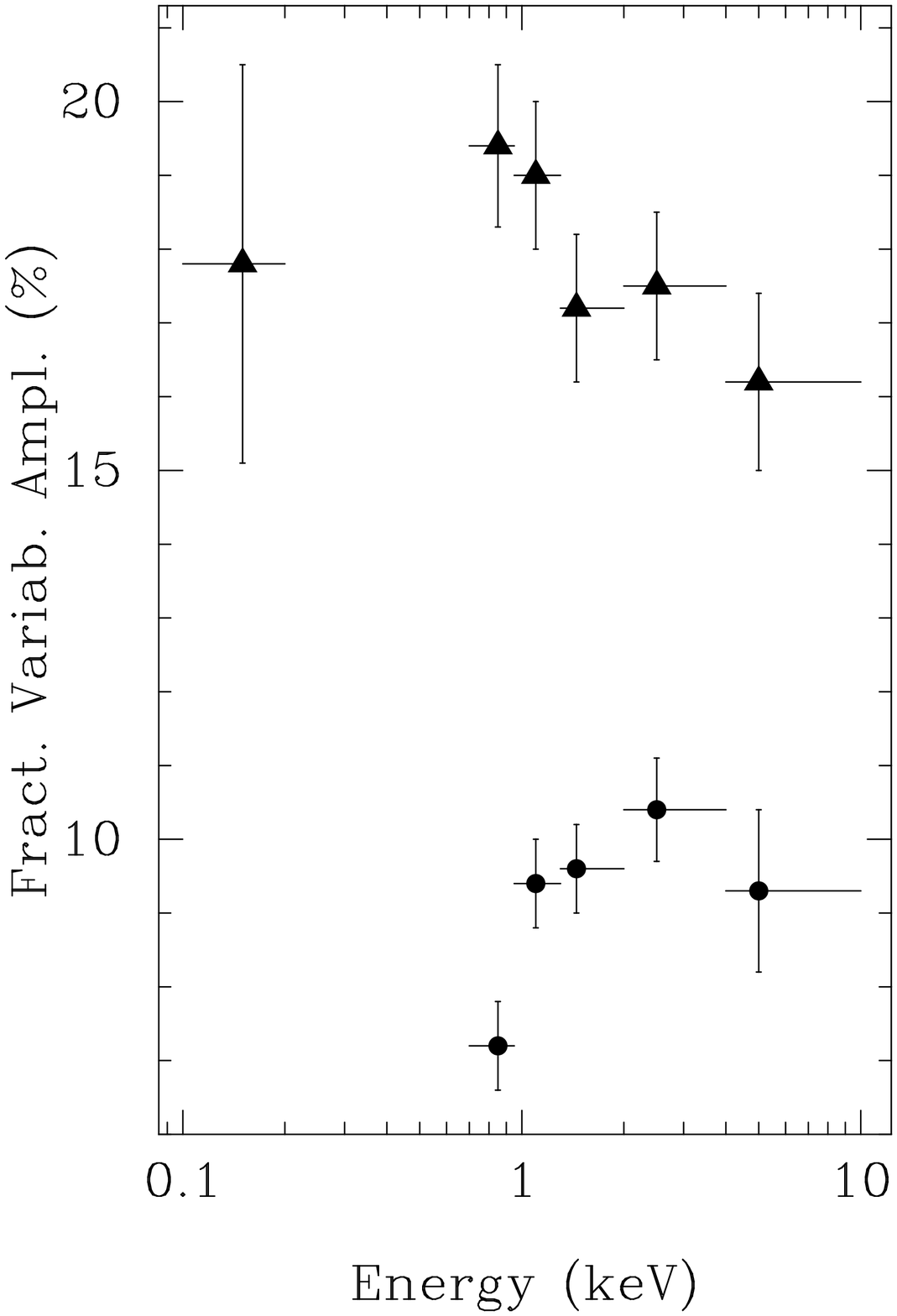}
\caption{
Fractional variability ($\fv$) and orbit-to-orbit variability amplitudes
($\fpp$) in different energy bands for \ark\ (left) and \tons\ (right).
The filled triangles refer to $\fv$, the fractional variability amplitude,
corrected for the effect of measurement noise, and the filled circles
refer to $\fpp$, the point-to-point variability amplitude, also corrected
for noise.
The error bars are derived as in the Appendix.
For the \euve\ observations of \tons, $\fpp$ is undefined (that is, the
measured variance is marginally smaller than that expected from
measurement noise alone), so no point is plotted.
}\label{fig4}
\end{figure}

\begin{figure}[ht]
\includegraphics{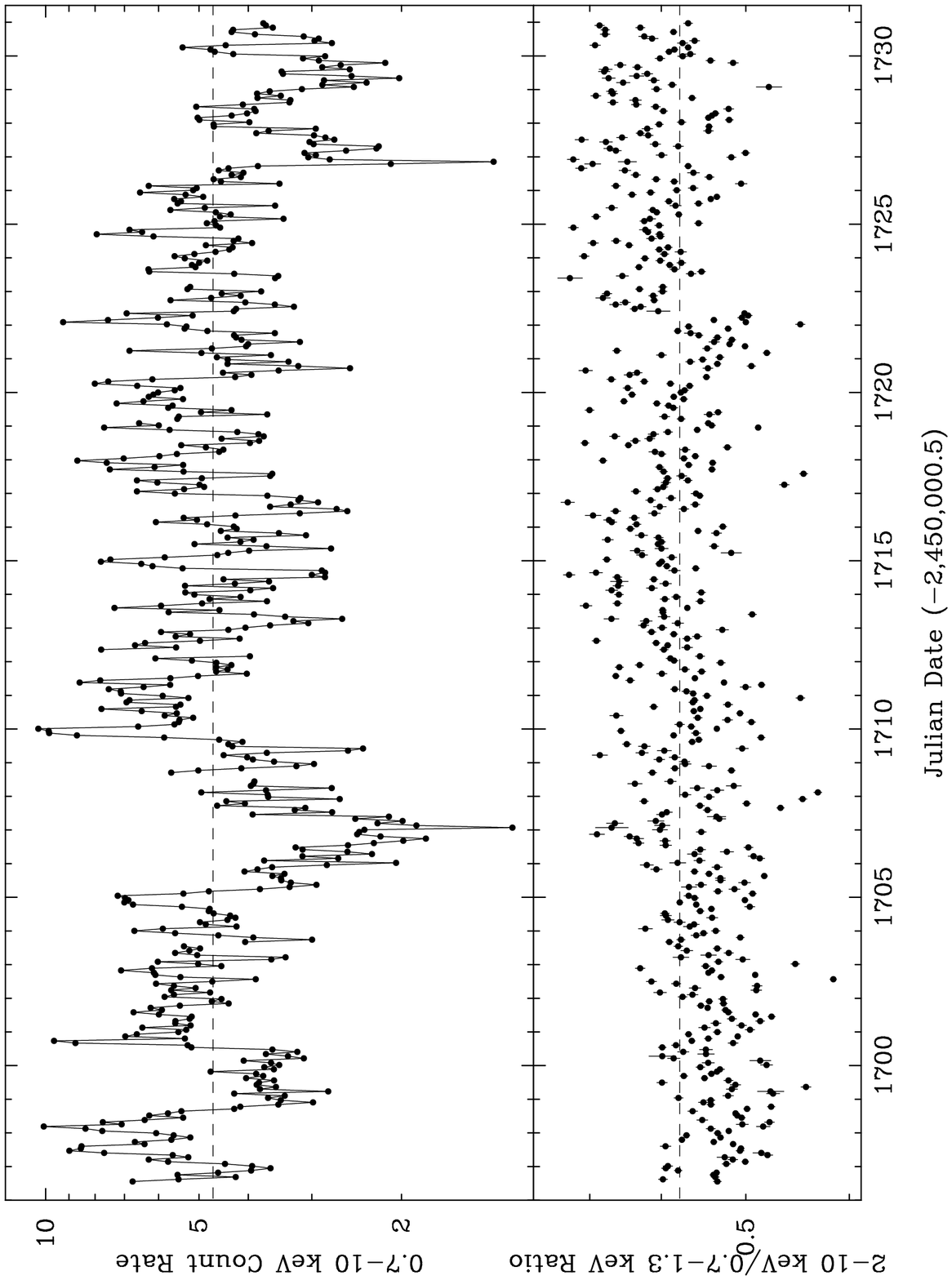}
\vspace{5.5in}
\caption{
Orbitally-binned flux and hardness ratio light curves for \ark.
The top panels show the 0.7--10~keV light curve while the
bottom panels show the 0.7--1.3~keV/2--10~keV hardness ratio.
The error bars are not shown on the fluxes; they are about the size of or
a bit bigger than the plotting symbols (typically $\sim$0.08~ct/sec).
}\label{fig5a}
\end{figure}

\setcounter{figure}{4}
\begin{figure}[ht]
\includegraphics{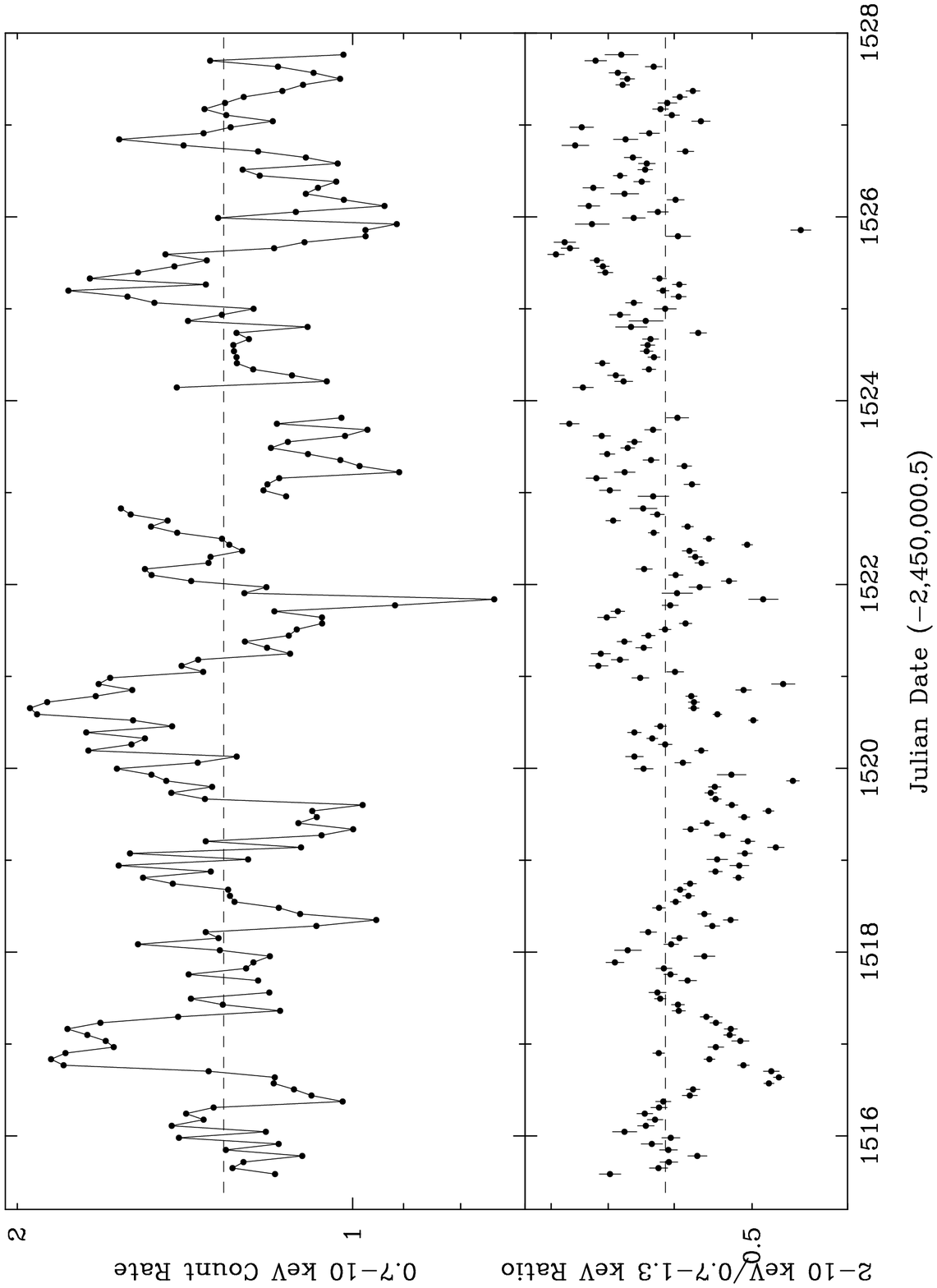}
\vspace{5.5in}
\caption{
Same as Figure~5a, but for \tons.
The error bars on the count rates are typically $\sim$0.03~ct/sec.
}\label{fig4b}
\end{figure}

\begin{figure}
\epsscale{0.85}
\plottwo{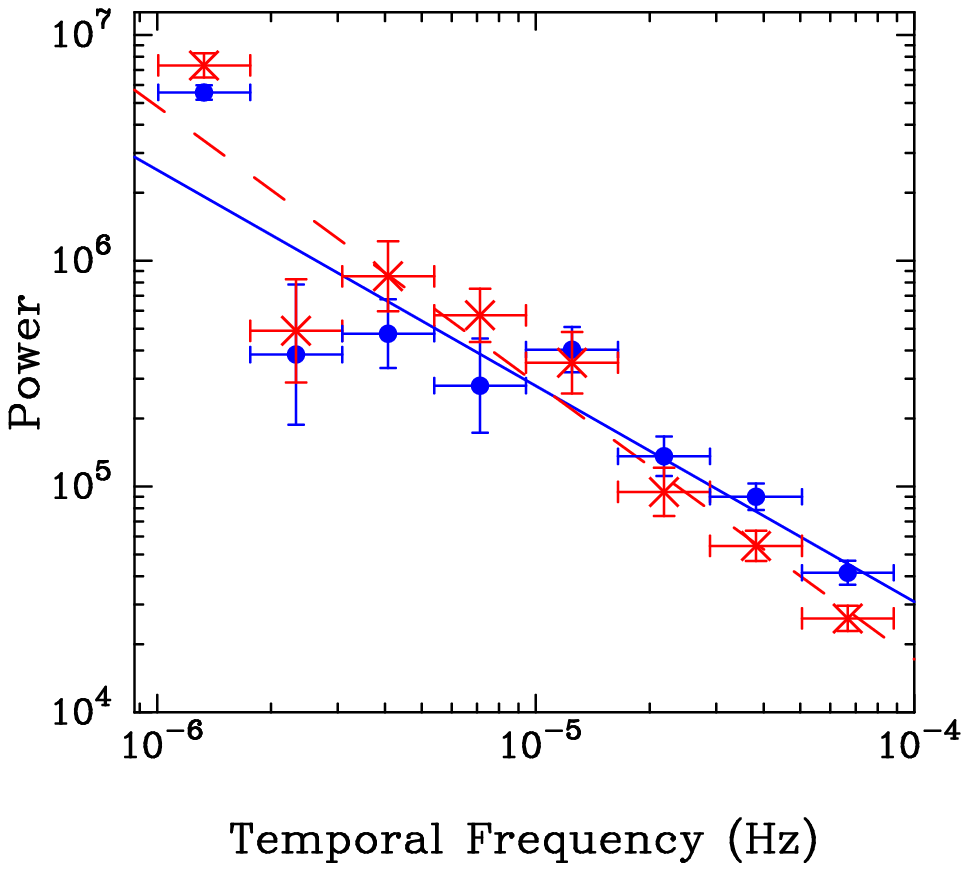}{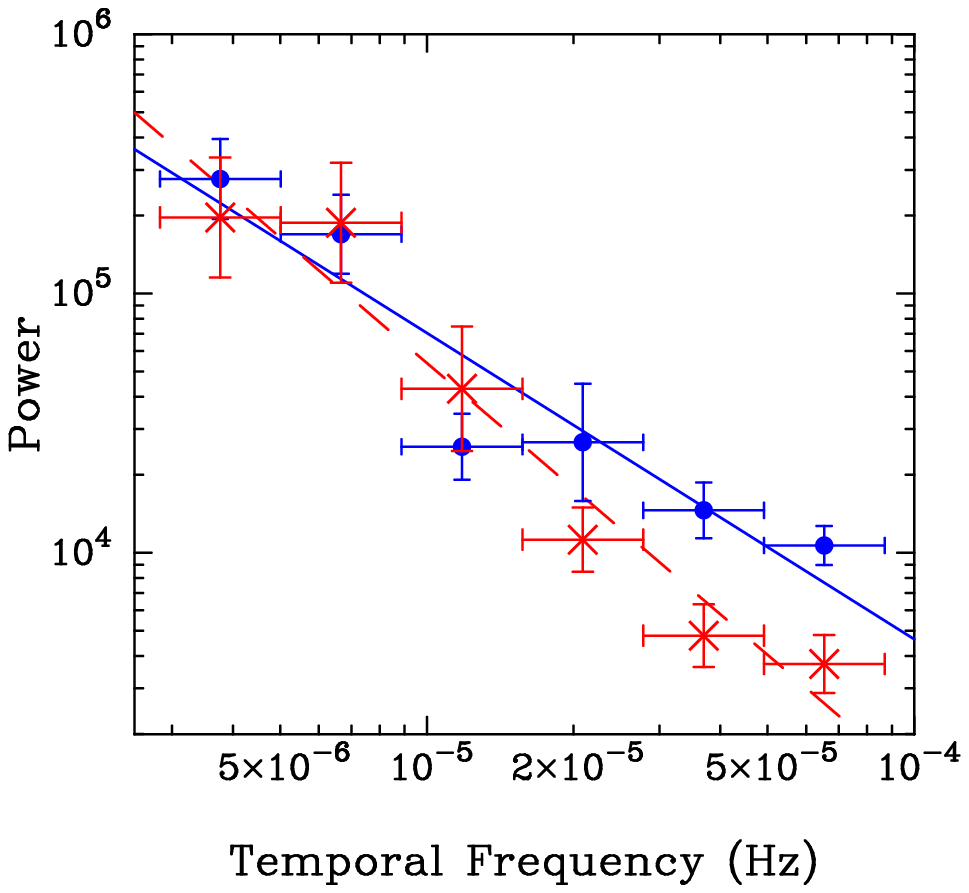}
\caption{
PDS for \ark\ (left) and \tons\ (right).
The 0.85~keV data are denoted in red by $\times$s and a dashed-line
power-law fit, while the 5~keV data are denoted in blue by circles and a
solid-line power-law fit.
}\label{fig6}
\end{figure}

\begin{figure}
\includegraphics{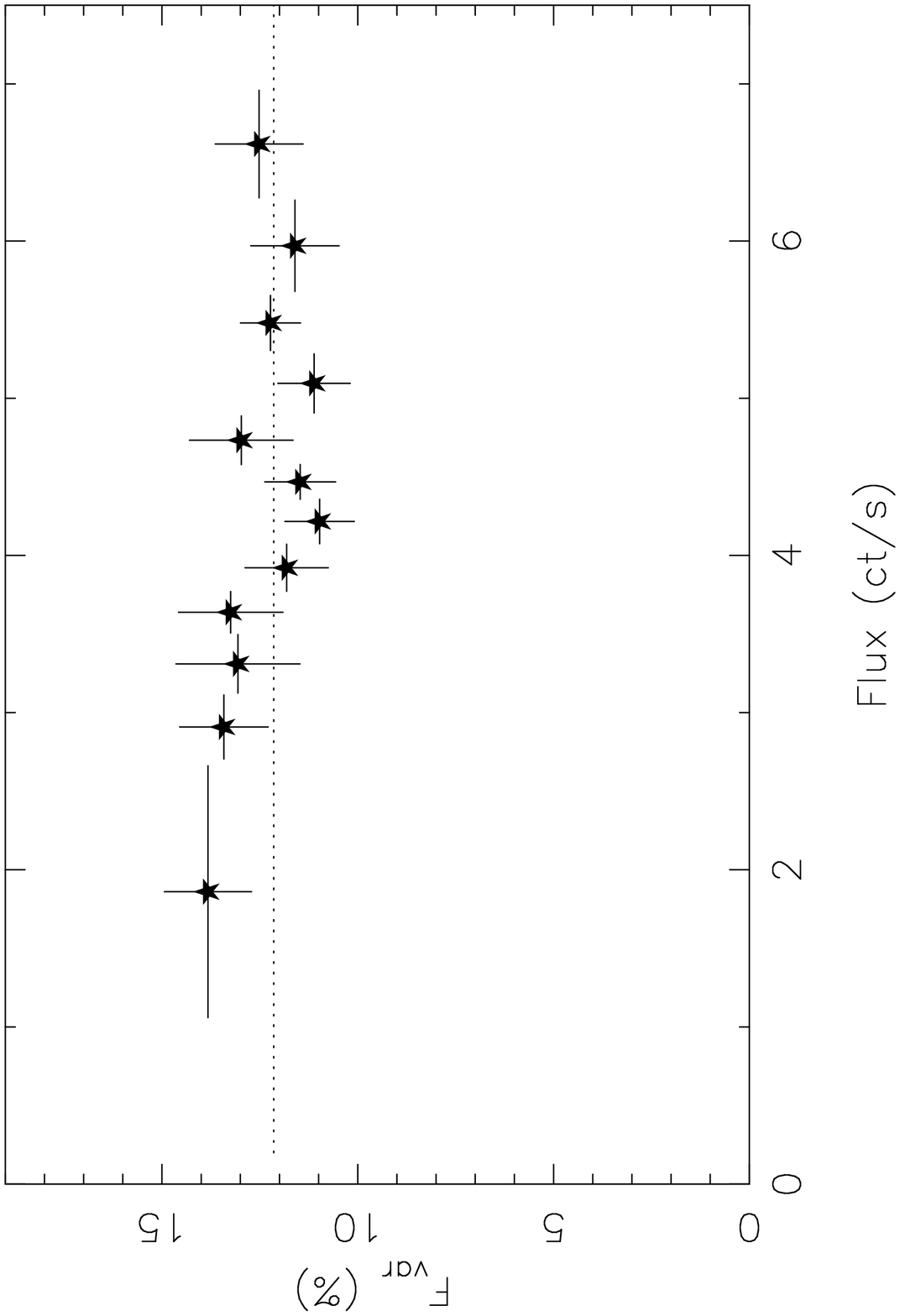}
\vspace{5.5in}
\caption{
Fractional variability amplitude binned as a function of mean count rate
for the 256 \ark\ orbits with more than 32 min of data.
As discussed in the text, the data were binned by flux such that each bin
has
at least 20 orbital points.
}\label{fig7}
\end{figure}

\begin{figure}[ht]
\includegraphics{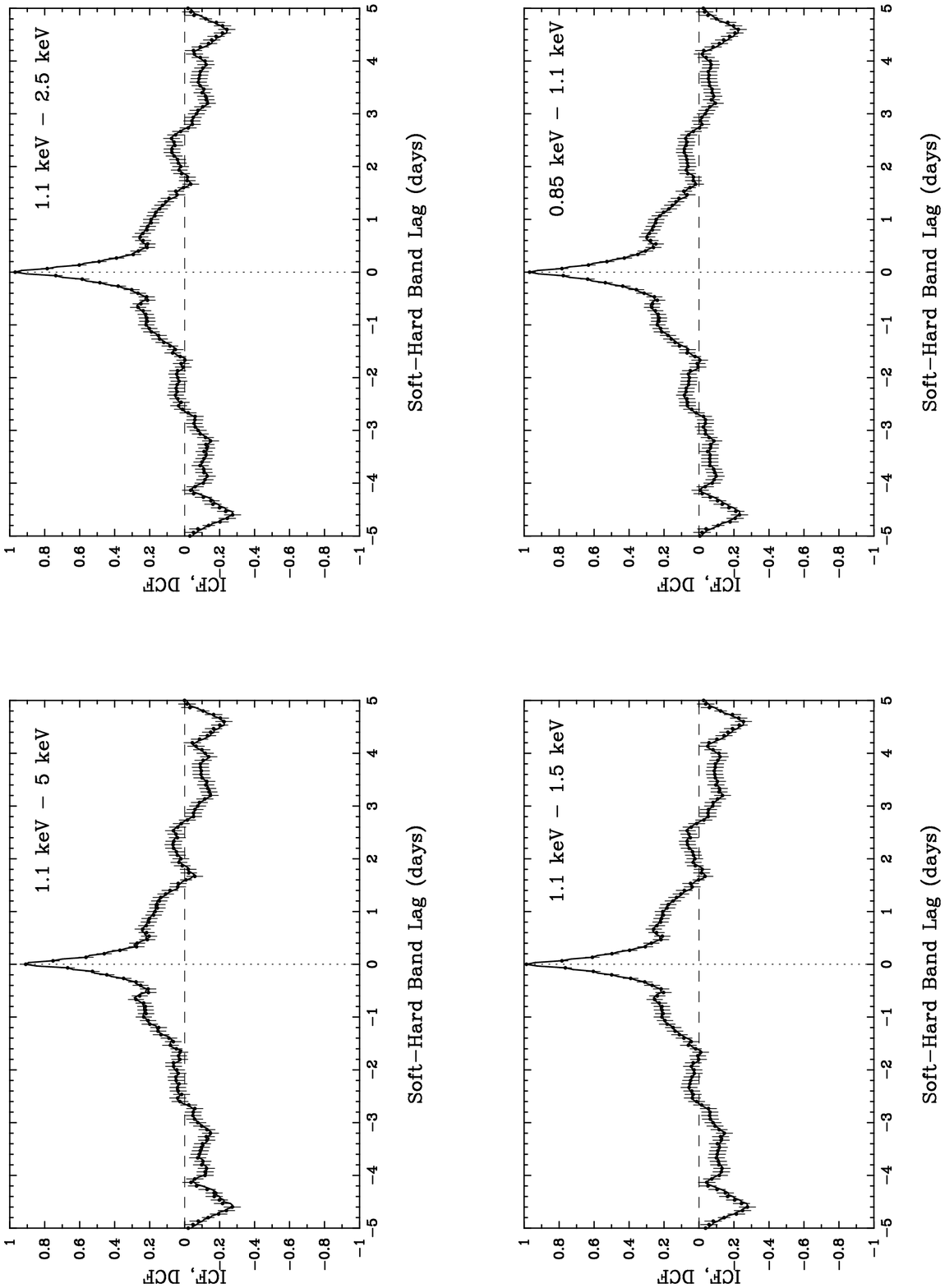}
\vspace{5.5in}
\caption{
Cross-correlation functions for \ark.
The solid line refers to the ICF, while the circles with error bars are
DCF.
All are referenced to 1.0~keV, in the sense that a positive peak would
mean that the softer band leads the harder.
The top panel is the CCF with 5~keV, next with 2.5~keV, next with 1.5~keV,
and at the bottom, with 0.7~keV.
}\label{fig8a}
\end{figure}

\setcounter{figure}{7}
\begin{figure}[ht]
\includegraphics{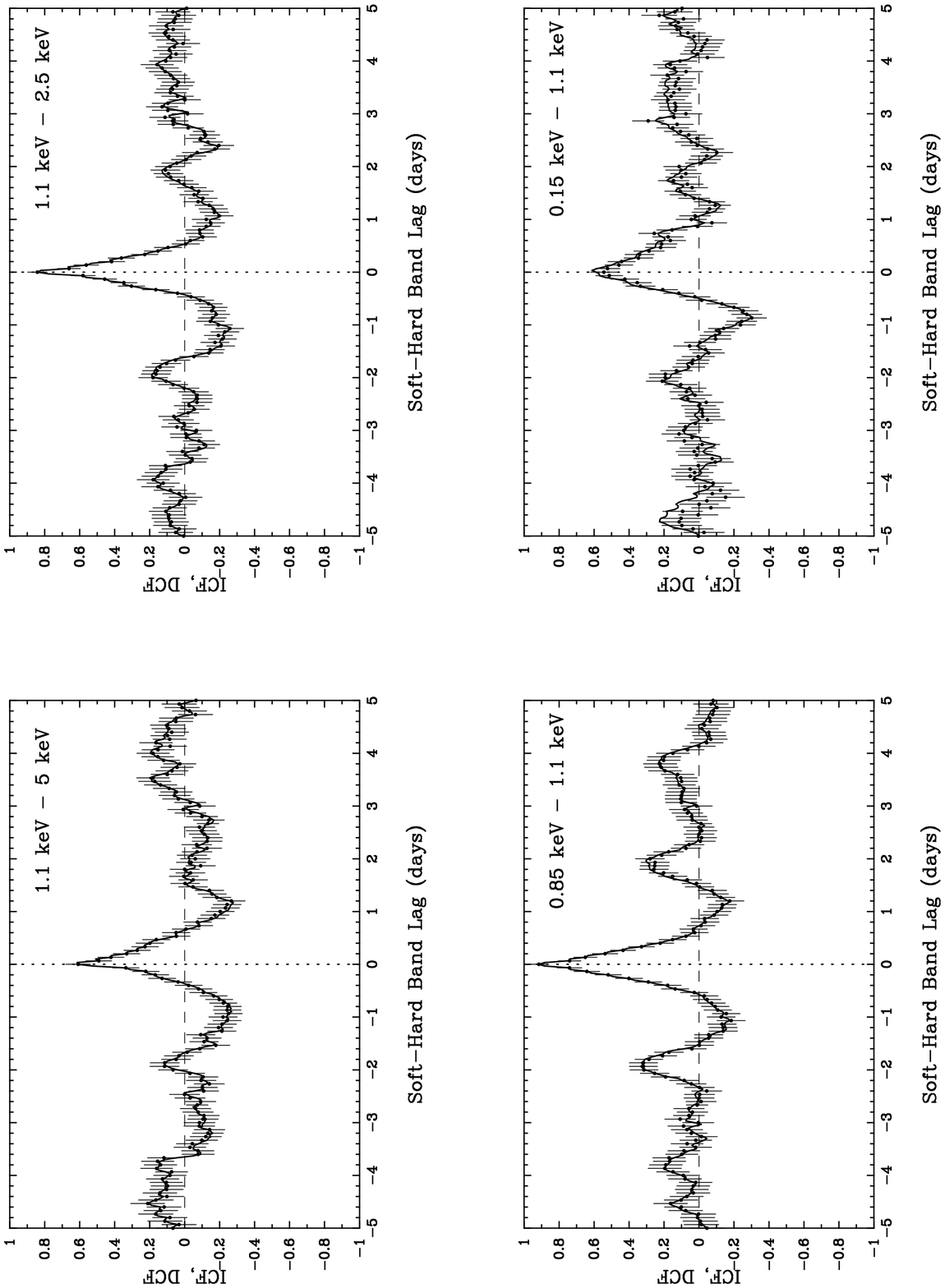}
\vspace{5.5in}
\caption{
Same as Figure~8a, except for \tons, and that the \asca\ 1.0~keV and
\euve\
0.15~keV band CCF is shown instead of the \asca\ 0.85~keV--1.5~keV CCF.
}\label{fig8b}
\end{figure}

\end{document}